\documentclass{emulateapj}

\journalinfo{To Appear in ApJ}
\slugcomment{Preprint version \today}

\shorttitle{The effective temperature scale of FGK stars. I.}

\shortauthors{Ram\'{\i}rez \& Mel\'endez}

\newcommand{\teff}{T_\mathrm{eff}}
\newcommand{\feh}{\mathrm{[Fe/H]}}
\newcommand{\logg}{\log g}
\newcommand{\fbol}{f_\mathrm{bol}}

\begin{document}

\title{The effective temperature scale of FGK stars. I. Determination of temperatures and angular diameters with the infrared flux method}

\author{Iv\'an Ram\'{\i}rez\altaffilmark{1}}
\affil{Department of Astronomy, University of Texas at Austin, RLM 15.306, TX 78712-1083} \and
\author{Jorge Mel\'endez\altaffilmark{1}} \affil{Department of Astronomy, California Institute of Technology, MC 105--24, Pasadena, CA 91125}

\altaffiltext{1}{Affiliated with the Seminario Permanente de Astronom\'{\i}a y Ciencias Espaciales of the Universidad Nacional Mayor de San Marcos, Peru}

\begin{abstract}
The infrared flux method (IRFM) has been applied to a sample of 135 dwarf and 36 giant stars covering the following regions of the atmospheric parameters space: 1) the metal-rich ($\feh\gtrsim0$) end (consisting mostly of planet-hosting stars), 2) the cool ($\teff\lesssim5000$ K) metal-poor ($-1\lesssim\feh\lesssim-3$) dwarf region, and 3) the very metal-poor ($\feh\lesssim-2.5$) end. These stars were especially selected to cover gaps in previous works on $\teff$ vs. color relations, particularly the IRFM $\teff$ scale of A.~Alonso and collaborators. Our IRFM implementation was largely based on the Alonso et al. study (absolute infrared flux calibration, bolometric flux calibration, etc.) with the aim of extending the ranges of applicability of their $\teff$ vs. color calibrations. In addition, in order to improve the internal accuracy of the IRFM $\teff$ scale, we recomputed the temperatures of almost all stars from the Alonso et al. work using updated input data. The updated temperatures do not significantly differ from the original ones, with few exceptions, leaving the $\teff$ scale of Alonso et~al. mostly unchanged. Including the stars with updated temperatures, a large sample of 580 dwarf and 470 giant stars (in the field and in clusters), which cover the ranges: 3600~K$\lesssim\teff\lesssim8000$~K, $-4.0\lesssim\feh\lesssim+0.5$, have $\teff$ homogeneously determined with the IRFM. The mean uncertainty of the temperatures derived is 75~K for dwarfs and 60~K for giants, which is about 1.3\% at solar temperature and 4500~K, respectively. It is shown that the IRFM temperatures are reliable in an absolute scale given the consistency of the angular diameters resulting from the IRFM with those measured by long-baseline interferometry, lunar occultation and transit observations. Using the measured angular diameters and bolometric fluxes, a comparison is made between IRFM and direct temperatures, which shows excellent agreement, with the mean difference being less than 10~K for giants and about 20~K for dwarf stars (the IRFM temperatures being larger in both cases). This result was obtained for giants in the ranges: 3800~K$<\teff<5000$~K, $-0.7<\feh<0.2$; and dwarfs in the ranges: 4000~K$<\teff<6500$~K, $-0.55<\feh<0.25$; and thus the zero point of the IRFM $\teff$ scale is essentially the absolute one (that derived from angular diameters and bolometric fluxes) within these limits. The influence of the bolometric flux calibration adopted is explored and it is shown that its effect on the $\teff$ scale, although systematic, is conservatively no larger than 50~K. Finally, a comparison with temperatures derived with other techniques is made. Agreement is found with the temperatures from Balmer line-profile fitting and the surface-brightness technique. The temperatures derived from the spectroscopic equilibrium of \ion{Fe}{1} lines are differentially consistent with the IRFM but a systematic difference of about 100~K and 65~K (the IRFM temperatures being lower) is observed in the metal-rich dwarf and metal-poor giant $\teff$ scales, respectively.
\end{abstract}


\section{Introduction} \label{sect:intro}

The effective temperature $(\teff)$ is one of the fundamental properties of a star. In principle, it may be determined from the stellar angular diameter $\theta$ and the bolometric flux $\fbol$ according to:
\begin{equation}
\fbol=\frac{\theta^2}{4}\sigma\teff^4\ , \label{eq:def1}
\end{equation}
where $\sigma$ is the Stefan-Boltzmann constant. In practice, however, the difficulties and limitations that arise when measuring angular diameters restrict the \emph{direct} determination of temperatures to a relatively low number of stars. The problem is worse for main sequence stars although interferometric (see e.g. Kervella et~al. 2004) and transit (Brown et~al. 2001) observations have been recently performed.

Even though there are alternative ways to determine effective temperatures, all of them require the introduction of models, which makes the results somewhat dependent on the models adopted. In the range of spectral types F0-K5, the effective temperature may be determined from the spectroscopic conditions of excitation and ionization equilibrium of Fe lines (e.g. Takeda et~al. 2002a, hereafter T02a), synthetic photometry (e.g. Bessell et~al. 1998, hereafter B98; Houdashelt et~al. 2000), fitting of the Balmer line-profiles (e.g. Fuhrmann 1998, hereafter F98; Barklem et~al. 2002), and line-depth ratios (e.g. Gray 1994; Kovtyukh et~al. 2003, hereafter K03). Our approach is the use of the infrared flux method (IRFM), which will be described in \S\ref{sect:irfm} and compared to the other methods in \S\ref{sect:IRFMcomparison}.

In purely spectroscopic methods, the temperature is determined by constraining its value in such a way that the abundance of \ion{Fe}{1} does not depend on the excitation potential of the lower level of the lines used and that the average abundance of Fe derived from \ion{Fe}{1} and \ion{Fe}{2} lines are equal (see e.g. T02a). This actually sets a constraint on both $\teff$ and $\logg$ (surface gravity) so degenerate solutions may be found. The technique has been applied to a wide variety of samples: stars in the thin and thick disks of the Galaxy (Bensby et~al. 2003, hereafter B03), planet-hosting stars (Santos et~al. 2004, hereafter SIM04), globular cluster stars (Sneden et~al. 2004), and so on. Evidence is being collected, however, towards the fact that the spectroscopic method has systematic errors caused by model uncertainties, such as non-LTE effects (note also that most spectroscopic studies use 1D-LTE classical model atmospheres). For instance, the spectroscopic temperatures are $\sim100$ K higher than those determined with the IRFM for planet-hosting stars (Ram\'{\i}rez \& Mel\'endez 2004b, hereafter RM04b; SIM04), surface gravities are $\sim0.1$ dex higher than those derived from Hipparcos parallaxes for $\feh>-0.7$ (Allende Prieto et~al. 1999, SIM04), and ionization equilibrium of Fe is not reproduced by classical model atmospheres neither in the Hyades cool dwarfs (Yong et~al. 2004) nor in the field metal-rich K dwarfs (Allende Prieto et~al. 2004, hereafter AP04).

Synthetic spectra and filter transmission functions have been used to generate grids of colors in the $\teff$, $\feh$, $\logg$ space using both Kurucz (1979) ATLAS models (B98) and Gustafsson et~al. (1975) MARCS models (Houdashelt et al. 2000). Synthetic colors are often put into the observational systems with samples of field stars (usually with metallicities near solar) with reliable temperatures or calibrated color-temperature relations. Zero point corrections are finally applied until the $\teff$ scale resembles the absolute one.\footnote{We refer to the $\teff$ scale obtained with direct temperatures (i.e. stellar angular diameter and bolometric flux measurements) as the \textit{absolute} $\teff$ scale.} Since this is made with field stars of solar metallicity, the zero point corrections are strictly valid only for synthetic colors with $\feh\sim0$, and its validity for other metallicities relies on the capability of models to reproduce real stellar spectra. Although observations and temperatures derived with other methods are involved in the procedure, the synthetic $\teff$ scales retain the basic properties of the model atmospheres adopted and are thus useful to test them.

The strength of the hydrogen Balmer lines is very sensitive to the effective temperature below 8000 K and it is only weakly affected by the surface gravity. Profile fitting is the best approach to an estimate of $\teff$ with Balmer lines, normally within 110 K of random error. Barklem et~al. (2002) found $5730\pm70$ K for the Sun with up-to-date line opacities and a reasonable agreement between their temperatures and those obtained with the IRFM for $\teff>5800$ K. Castelli et~al. (1997), however, show that the results from Balmer line-profile fitting are strongly dependent on the models and the details of the convection treatment. To satisfactorily reproduce the profiles in the solar spectrum using ATLAS9 models with overshoot, for example, the temperature of the Sun has to be increased up to 5900 K.

With the line-depth ratio technique (Gray \& Johansson 1991, Gray 1994) very small internal errors ($\sim10$~K) for individual stars may be achieved if several calibrated line-depth ratios are used. Even though the temperature scale derived with this technique may be less accurate due to systematic errors, the line-depth ratios may be used to look for very small time dependent $\teff$ variations, like those produced by photospheric inhomogeneities and starspots. The solar temperature with this method results to be 44~K below the standard value of 5777~K and an empirical correction of 44~K has to be applied (K03). Since nothing guarantees that this correction is applicable to the whole range of atmospheric parameters, the systematics, if present, are not removed.

The aim of this work is to extend and improve the internal accuracy of the best available empirical temperature-color calibrations based on the IRFM, namely, those given by Alonso et~al. (1996b, hereafter AAM96b; 1999b, hereafter AAM99b)\footnote{Hereafter, the series of papers by A. Alonso and collaborators are referred to as AAM.} for the $(UBVRI)_\mathrm{J}(JHK)_\mathrm{TCS}$ and Str\"omgren $uvby$-$\beta$ systems. Mel\'endez \& Ram\'{\i}rez (2003) and Ram\'{\i}rez and Mel\'endez (2004a, hereafter RM04a) extensions to the Vilnius, Geneva, $RI_\mathrm{(C)}$ and DDO systems will also be revised. In this paper we describe our IRFM implementation and test our results. The calibration of the metallicity-dependent $\teff$ vs. color relations is the subject of a companion paper (hereafter referred to as Part II).

This paper is distributed as follows: in \S\ref{sect:irfm} the IRFM and its present implementation are explained, the sample and input data adopted as well as their effects on the temperatures to be derived are described in \S\ref{sect:sample}. The calculation of the final temperatures and angular diameters is given in \S\ref{sect:IRFMtemps} while the fundamental tests (e.g. comparison with measured angular diameters) and comparison with other methods are presented in \S\ref{sect:IRFMcomparison}. Conclusions are summarized in \S\ref{sect:conclusions}.

\section{The IRFM and its present implementation} \label{sect:irfm}

The IRFM was first introduced by Blackwell \& Shallis (1977) as a mean to simultaneously determine the effective temperature and angular diameter of a star. The basic idea of the method is to use the emergent monochromatic flux predicted by models ($\psi_\lambda$) and, similarly to Eq.~(\ref{eq:def1}), given by
\begin{equation}
f_\lambda=\frac{\theta^2}{4}\psi_\lambda(\teff,\feh,\logg)\ ,
\label{eq:def2}
\end{equation}
where $f_\lambda$ is the corresponding flux as measured on Earth, to solve the system of equations (\ref{eq:def1}) and (\ref{eq:def2}) for the unknowns $\teff$ and $\theta$. The IRFM thus requires $\fbol$, $\feh$, $\logg$ and $f_\lambda$ to be known \textit{a priori}. The use of the infrared is due to the relatively low dependence of $\psi_\lambda$ on $\teff$ in this region of the spectrum (see e.g. Fig. 1 in Blackwell et~al. 1979).

A more practical approach to the IRFM was given by Blackwell et~al. (1980), who defined the $R$-factors as the quotients between the total and monochromatic fluxes. The
comparison of the observational $R$-factor ($R_\mathrm{obs}=\fbol/f_\lambda$) with a grid of theoretical $R$-factors ($R_\mathrm{theo}=\sigma\teff^4/\psi_\lambda$, note that $R_\mathrm{theo}$ is a grid in the $\teff$, $\feh$, $\logg$ space at a given wavelength) allows to easily automatize the method.

The IRFM is often applied using more than one infrared wavelength ($\lambda$), and a different $\teff$ is obtained for each $\lambda$. A critical ingredient of the IRFM is the absolute flux calibration of the standard star from which the $f_\lambda$ fluxes of the problem stars are calculated. If $f_\lambda^\mathrm{cal}$ are the absolute monochromatic fluxes for the standard star, then
\begin{equation}
f_\lambda=q_\lambda\times
f_\lambda^\mathrm{cal}10^{-0.4(m-m_\mathrm{cal})}\ ,
\label{eq:fluxes}
\end{equation}
where $m$ is the observed magnitude in a given band for which $\lambda$ is, for example, the mean wavelength of the filter. The $q_\lambda$ ($\sim1$) factors are required since the filters have
finite widths and are not usually narrow in the infrared. It would be ideal to calculate the $q_\lambda$ factors from spectrophotometric observations but in practice models are adopted (see the appendix in AAM94a for a discussion on this particular point).

For the IRFM, the absolute flux calibration for the standard star must be done in such a way that the temperatures obtained in different wavelengths are close to each other and, when properly
averaged, consistent with the direct temperatures. A calibration of this kind is that of AAM94a and is the one we have used in our work. Even though it was derived using direct temperatures of
giant stars, this calibration provides good temperatures for dwarf stars as well, as it was shown in RM04b, where the comparison of IRFM angular diameters with measured ones for 7 main sequence stars showed good agreement (see also \S\ref{sect:IRFMvsDIAM} for an update).

\begin{figure*}
\epsscale{0.8}
\plotone{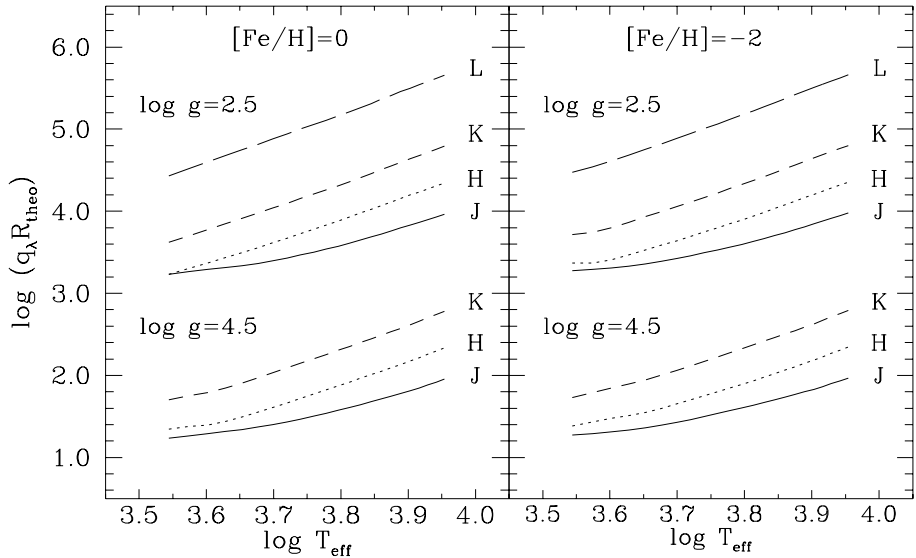} \caption{Theoretical $qR$ factors adopted as a function of effective temperature for typical atmospheric parameters of dwarf ($\logg=4.5$), giant ($\logg=2.5$), metal-rich ($\feh=0.0$), and metal-poor ($\feh=-2.0$) stars; and filter. The units of $R_\mathrm{theo}$ are \AA\ for $\logg=2.5$. For $\logg=4.5$ the lines have been shifted by $-2.0$ dex. A rough estimate leads to $qR\propto\teff^{2.7}$, almost independently of the band.} \label{fig:qr}
\end{figure*}

The models we have adopted for the $\psi_\lambda$ fluxes are Kurucz SEDs (Spectral Energy Distributions) as given in the library of theoretical spectra by Lejeune et~al. (1997). From them and the response functions of the $J$, $H$, $K$ and $L$ filters of the TCS system (AAM94b, AAM98) we calculated grids of $q_\lambda$ and $R_\mathrm{theo}$ factors
in the ranges 3500~K$<\teff<9000$~K, $-3.5<\feh<+0.5$, and $0.0<\logg<5.0$, where Kurucz models are available. Figure \ref{fig:qr} illustrates the dependence of the temperature indicator in the IRFM (the $q_\lambda\times R_\mathrm{theo}$ factors, see \S\ref{sect:IRFMtemps}) as a function of atmospheric parameters and filter.

In the ranges of most interest to us ($\teff>3800$~K for giants, $\teff>4500$~K for dwarfs), the use of more sophisticated models will not have a substantial effect on the IRFM given that the model dependence of the method is relatively small. Also, unlike the UV and optical regions of the spectrum, the density of lines in the infrared is relatively low. The SEDs we took from Lejeune et~al. (1997) are originally from Kurucz (1992). They were synthesized with Kurucz multi-million line list and atmosphere models with scaled solar abundances and an approximate treatment of convective overshooting. The models were computed with a microturbulent velocity of 2~km~s$^{-1}$.

The use of classical model atmospheres may not be a big issue for the IRFM. It is remarkable that the three-dimensional hydrodynamical simulations suggest that the effective temperature scale derived with the IRFM using one-dimensional hydrostatic models is the same as the one obtained with hydrodynamical models for turn-off stars with $-3<\feh<0$, and that for stars in the same metallicity range and approximately solar temperature the difference is only about 20~K (Asplund \& Garc\'{\i}a P\'erez 2001).

\section{Sample, photometry and atmospheric parameters adopted} \label{sect:sample}

Our sample consists of stars in the ranges of atmospheric parameters for which few data points were available in AAM work, like the metal-poor and metal-rich extremes of the regions around $\teff\sim5800$~K and below 5000~K. The planet-hosting star sample is a metal-rich one given the well stated metallicity-enhancement of stars with planets (see e.g. Gonzalez et~al. 2001, SIM04). Other metal-rich stars were taken from Castro et~al. (1997) and Pompeia et~al. (2002). The very metal-poor stars added are mainly from Ryan et~al. (1991, 1996), Norris et~al. (2001, mostly hot FG unevolved stars), and McWilliam et~al. (1995); while the metal-poor cool dwarfs are from Yong \& Lambert (2003a, 2003b).

\begin{figure}
\epsscale{1.1}
\plotone{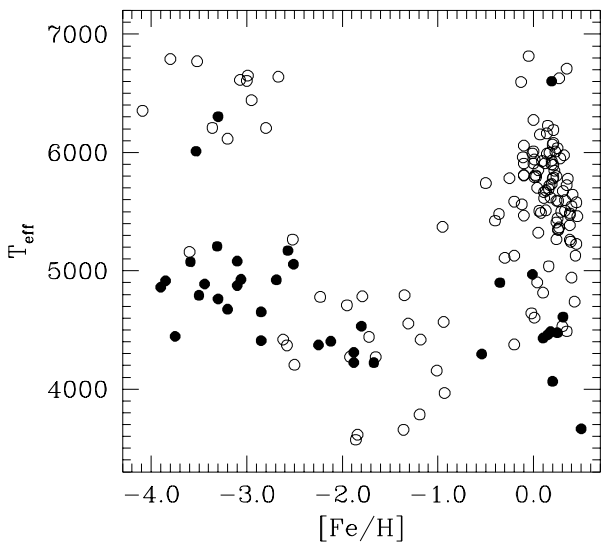} \caption{Distribution of our sample stars in the $\teff$ vs. $\feh$ plane, open circles: dwarf stars; filled circles: giant stars.} \label{fig:NewSample}
\end{figure}

The distribution of our sample stars in the $\teff$ vs. $\feh$ plane is given in Fig. \ref{fig:NewSample}.

\begin{figure*}
\epsscale{1.1}
\plotone{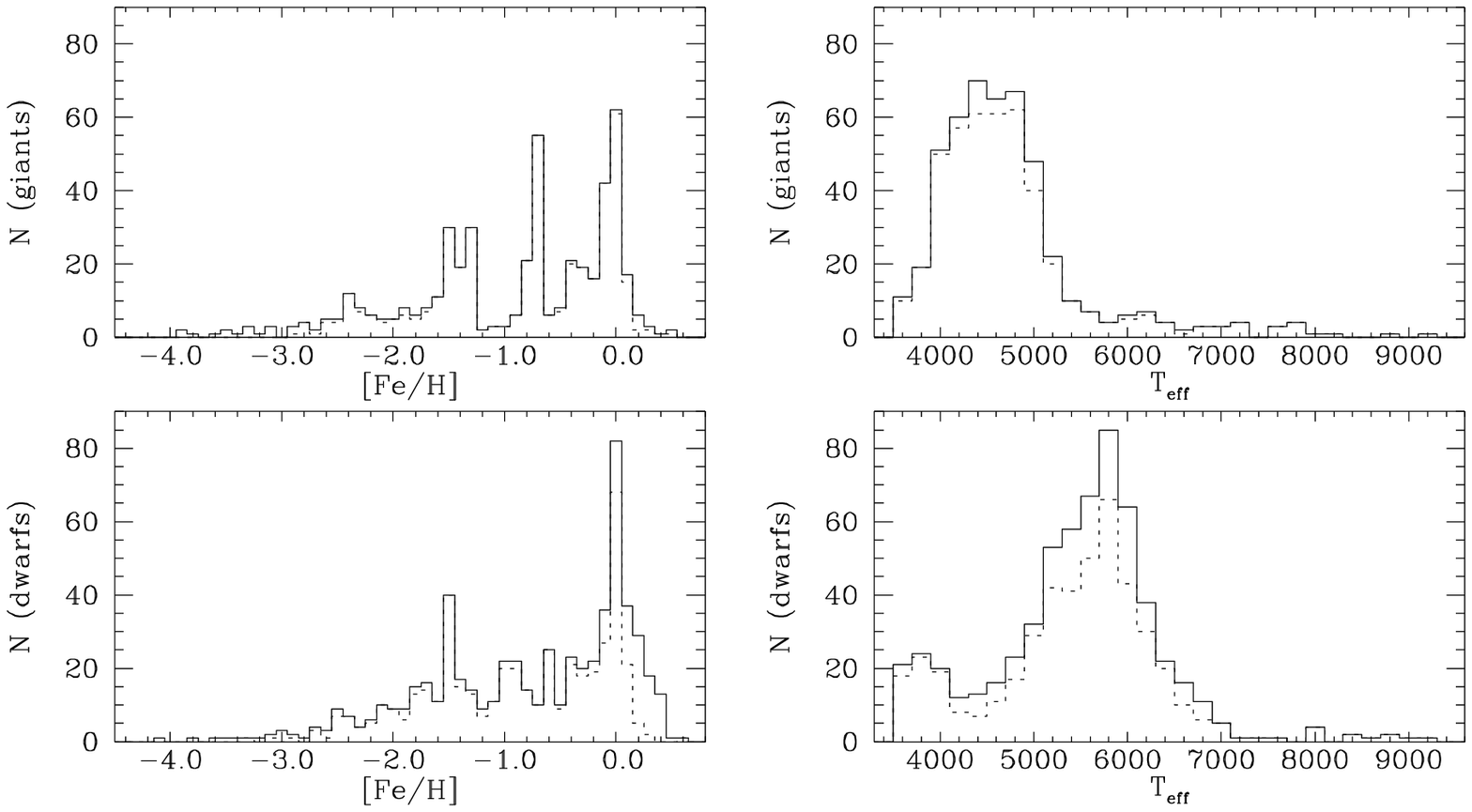} \caption{Distribution of the AAM (dotted lines) and combined (AAM and this work, solid lines) giant (top panels) and dwarf (bottom panels) samples in $\feh$ and $\teff$. $N$ is the number of stars in 0.1 dex bins in $\feh$ or 200 K bins in $\teff$.} \label{fig:sample}
\end{figure*}

The distribution of the combined sample (i.e. including AAM stars) in $\teff$ and $\feh$ is shown in Fig.~\ref{fig:sample}. The total sample is shown with the solid histogram, the dotted histogram is the AAM sample. Some of the peaks in the giant $\feh$ distribution correspond to giants in clusters included in AAM work while those at $\feh=0.0,-1.0,$ and $-1.5$ in the dwarf $\feh$ distribution are the stars with kinematic metallicity assignment (see \S\ref{sect:feh}). Note how the dwarf metal-rich extreme is now much more populated, mainly with the planet-hosting stars. The very metal-poor end of both dwarfs and giants is also better represented, although still in small numbers. Some hot stars ($\teff>7500$~K) are also included but they are almost in the limit of applicability of the IRFM.

\subsection{Photometry and reddening corrections}

The infrared photometry for the AAM sample was taken from AAM94b and AAM98. This photometry is given in the TCS system and has a mean accuracy of 0.02 mag. For our sample, Johnson and 2MASS photometry was adopted and transformed into the TCS system by means of the transformation equations of AAM94b and AAM98 (Johnson to TCS) and the following transformation equations (2MASS to TCS) we derived for both dwarfs:
\begin{mathletters}
\begin{eqnarray}
J_\mathrm{TCS} & = & J_\mathrm{2MASS}+0.001-0.049(J-K)_\mathrm{2MASS}\ ,\\
H_\mathrm{TCS} & = & H_\mathrm{2MASS}-0.018+0.003(J-K)_\mathrm{2MASS}\ ,\\
K_\mathrm{TCS} & = &
K_\mathrm{2MASS}-0.014+0.034(J-K)_\mathrm{2MASS}\ ,
\end{eqnarray}
\end{mathletters}
with standard deviations of $\sigma=0.038$ ($N=104$),
$\sigma=0.039$ ($N=103$) and $\sigma=0.035$ ($N=107$),
respectively ($N$ is the number of stars in every fit); and
giants:
\begin{mathletters}
\begin{eqnarray}
J_\mathrm{TCS} & = & J_\mathrm{2MASS}+0.012-0.063(J-K)_\mathrm{2MASS}\ ,\\
H_\mathrm{TCS} & = & H_\mathrm{2MASS}-0.013-0.009(J-K)_\mathrm{2MASS}\ ,\\
K_\mathrm{TCS} & = &
K_\mathrm{2MASS}-0.014+0.027(J-K)_\mathrm{2MASS}\ ,
\end{eqnarray}
\end{mathletters}
with standard deviations of $\sigma=0.036$ ($N=43$),
$\sigma=0.038$ ($N=40$) and $\sigma=0.027$ ($N=43$), respectively.

The $V$ magnitudes were mostly obtained from the different catalogs included in the General Catalogue of Photometric Data (Mermilliod et~al. 1997, hereafter GCPD).

For the globular and open clusters we adopted the reddening corrections given by Kraft \& Ivans (2003) and AAM99a, respectively. For the field stars in AAM sample, their $E(B-V)$ values were adopted, while for our sample stars the reddening was estimated as explained below.

We estimated $E(B-V)$ values from the maps of Schlegel et al. (1998) and Burstein \& Heiles (1982), several extinction surveys included in Hakkila et~al. (1997) code, and empirical laws by Bond (1980) and Chen et~al. (1998). For stars closer than 75 pc, $E(B-V)=0$ was adopted. The Arenou et~al. (1992) extinction model included in Hakkila et~al. code seems to systematically overestimate the reddening of stars with $d<0.5$ kpc, and so we have given a lower weight to the $E(B-V)$ values obtained from Arenou et~al. map for stars closer than 500 pc. A latitude-dependent weight was adopted for the reddening obtained from the Schlegel et~al. and Burstein \& Heiles maps, in such a way that the high latitude objects ($|l|>50$) had a larger weight and the low latitude objects ($|l|<30$) a weight that was close to zero.

There is evidence that the $E(B-V)$ values obtained from the Schlegel et~al. map, which is based on dust maps by DIRBE/COBE observations, overestimates the reddening, so we have multiplied the Schlegel et~al. $E(B-V)$ values by 0.8 (Arce \& Goodman 1999; Chen et~al. 1999; Beers et~al. 2002; Dutra et~al. 2002, 2003). 

The extinction ratios $k=E(\mathrm{color})/E(B-V)$, or $k=A_\mathrm{band}/E(B-V)$ given in Table~\ref{t:red} were obtained from Schlegel et~al. (1998), using the appropriate effective wavelengths of the filters.

\begin{deluxetable}{rr}
\tablecaption{Extinction Ratios Adopted, $k=E(\mathrm{color})/E(B-V)$ or $A_\mathrm{band}/E(B-V)$.}
\tablehead{\colhead{Color or band} & $k$ (mag)}
\startdata
$(V-K_\mathrm{TCS})$ & 2.74 \\
$(V-K_\mathrm{J})$ & 2.72 \\
$J_\mathrm{TCS}$ & 0.84 \\
$H_\mathrm{TCS}$ & 0.56 \\
$K_\mathrm{TCS}$ & 0.36 \\
$K_\mathrm{J}$ & 0.38 \\
$L_\mathrm{TCS}$ & 0.15
\enddata
\label{t:red}
\end{deluxetable}

\subsection{Metallicities} \label{sect:feh}

Our basic reference for the $\feh$ values is the Cayrel de Strobel et~al. (2001) catalog. We have updated the catalog with recent spectroscopic determinations, the most important of them being Mishenina \& Kovtyukh (2001), Santos et~al. (2001), Heiter \& Luck (2003), Stephens \& Boesgaard (2002), Takada-Hiday et~al. (2002), and Yong \& Lambert (2003a). From now on, we will refer to this updated catalog as C03.

For the stars with as many or more than four entries in the catalog, the standard deviation of the different measurements was adopted as the error bar for $\feh$, which was assumed to be the average of the values reported in C03. For those with less than four entries a mean error of 0.15 dex was adopted.

Based on the C03 catalog and Str\"omgren photometry, we were able to calculate the metallicities of $\sim$20\% of the sample dwarf stars from the following photometric calibration: 

\noindent For $0.19\leq(b-y)<0.35$, with $\sigma=0.17$ dex:
\begin{eqnarray}
\feh & = & -4.29-66.0m_1+444.2m_1(b-y) \nonumber \\
& & -782.4m_1(b-y)^2  \\
& & +(0.966-37.8m_1-1.707c_1)\log\eta\ , \nonumber
\end{eqnarray}
where $\eta=m_1-[0.40-3.0(b-y)+5.6(b-y)^2]$.

\noindent For $0.35\leq(b-y)<0.50$, with $\sigma=0.13$ dex:
\begin{eqnarray}
\feh & = & -3.864+48.6m_1-108.5m_1^2 \nonumber\\
& & -85.2m_1(b-y)+190.6m_1^2(b-y) \\
& & +[15.7m_1-11.1c_1+17.7(b-y)]c_1\ .\nonumber
\end{eqnarray}

\noindent For $0.50\leq(b-y)_0\leq0.80$, with $\sigma=0.15$ dex:
\begin{eqnarray}
\feh & = & -2.63+26.0m_1-41.3m_1^2-45.4m_1(b-y) \nonumber \\
& & +74.0m_1^2(b-y)+17.0m_1c_1\ .
\end{eqnarray}

A mean error of 0.2 dex was adopted for these photometric metallicities. These formulae follow Schuster \& Nissen (1989) parameterization, but our relations are based on a much larger number of stars ($\sim800$) and they extend to late K stars. The formulae are valid for $\logg>3.4$ and $-2.5<\feh<0.4$. For F dwarfs they extend to $\feh\sim-3.5$. The major part of the dispersion comes from the metal-poor stars, for $\feh>-1.5$ the standard deviations reduce to 0.1 dex.

The metallicities derived by SIM04 were found consistent with C03 and were thus adopted for the sample of planet-hosting stars. Although the temperatures from the IRFM and those derived from a purely spectroscopic approach differ by 100 K the metallicities are not severely affected if another $\teff$ scale is adopted (J.~Mel\'endez \& I.~Ram\'{\i}rez 2005, in preparation).

The photometric metallicities derived by AAM96a and AAM99a were also adopted, when necessary, and a mean error of 0.25 dex was assumed. Kinematic metallicities were assigned as a last resource. For these stars ($\sim10$\% of the dwarf and field giant samples) a mean error of 0.5~dex in $\feh$ was adopted. The criteria was $\feh=0.0$ for thin disk stars, $\feh=-1.0$ for thick disk stars, and $\feh=-1.5$ for halo stars. The temperatures for these stars are very unreliable and should be taken only as rough estimates.

Most of the metallicites for the giant stars in clusters have been taken from Kraft \& Ivans (2003). For the clusters not included in Kraft \& Ivans work, AAM99a values were adopted (see AAM99a for the references).

\subsection{Surface gravities}

As we did with the metallicities, we adopted the mean values from the C03 catalog when as many or more than four entries were available. These include both spectroscopic and Hipparcos gravities and even though it has been suggested that a systematic difference of the order of 0.1 dex exists between them, it has almost no effect on the $\teff$ derived with the IRFM (see e.g. Fig. 4 in AAM96a). For the stars with less than four entries a mean error of 0.30 dex was adopted.

The exact value of $\logg$ is not required for the IRFM, a distinction between giant and dwarf is, in principle, reasonable. Thus, when no spectroscopic or Hipparcos gravities were available, the $\logg$ value was estimated from the spectral type and luminosity class as given in Simbad and the representative $\logg$ values given by Strai\v{z}ys (1995). An error bar of 0.6 dex was assumed in this case. The $\logg$ values of AAM96a and AAM99a were used for the remaining stars (0.7 dex of mean error).

The surface gravities of the cluster giants were calculated from their distance moduli (Kraft \& Ivans 2003) and the bolometric corrections of AAM99a, taking $0.85M_\odot$ as an estimate of the stellar masses.

\subsection{Bolometric fluxes}

Bolometric fluxes have been calculated from the $K$ magnitudes and $(V-K)$ colors, using the calibrations of AAM95 and AAM99a, which are based on integrated $UBVRIJHK$ photometry. These calibrations have a mean accuracy of 1\% and are in reasonable agreement with other flux determinations (AAM99a, RM04b).

\begin{figure}
\epsscale{1.15}
\plotone{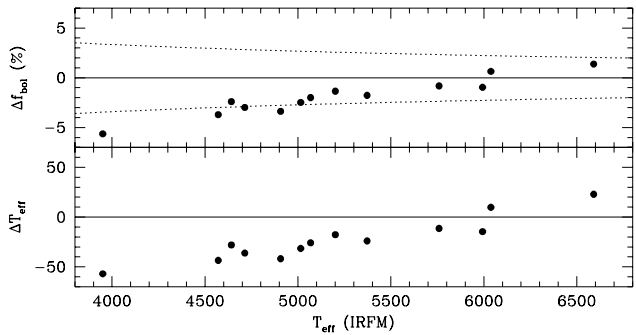} \caption{Comparison of AAM95 bolometric flux calibration with B98 theoretical BC scale and their effect on the $\teff$ scale. Upper panel: fluxes from the AAM95 calibration minus B98. The dotted lines show how the fluxes would need to be off as to produce an IRFM $\teff$ scale shifted by $\pm$50 K. Lower panel: our IRFM temperatures minus B98 (see the text for details).} \label{fig:a95b98}
\end{figure}

Systematically inaccurate fluxes lead to offsets and/or trends in the temperature scale. As an illustration, we have calculated self-consistent bolometric fluxes and temperatures from the theoretical bolometric correction (BC) scale of B98 and compared them to ours for 13 dwarf stars\footnote{These 13 stars are used in \S \ref{sect:IRFMvsDIAM} to compare measured diameters and direct temperatures with the IRFM (see Table \ref{t:treceD}).} in Fig. \ref{fig:a95b98}. We derived $\fbol$ values from their tables of BC by calculating an initial flux $\fbol(1)$ from our IRFM $\teff$ and the adopted parameters. Then, using the $\fbol$ from the calibration of AAM95 and $\fbol(1)$ we got $\teff(2)=\teff(\fbol(1)/\fbol)^{1/4}$ and iterated. This procedure is safe, because it converges simultaneously to $\fbol$ and $\teff$  independently of the input values, which is not true if we use the measured angular diameters to iterate in $\teff$.

Figure \ref{fig:a95b98} shows how different bolometric flux scales lead to different temperature scales. Lower temperatures are clearly correlated to lower fluxes, although the correlation is not linear. While the difference in the fluxes increases by 7\% (from $\sim-5.5$\% at $\teff\sim4000$ K to $\sim1.5$\% at $\teff\sim6500$~K), the difference in the temperature scales increases only by about 2\% (from $-60$ K or $\sim-1.5$\% at $\teff\sim4000$ K to 20 K or $\sim0.5$\% at $\teff\sim6500$ K ). Also shown in Fig. \ref{fig:a95b98} as dotted lines are the flux differences that would produce a $\teff$ scale offset of $\pm$50 K in the IRFM.\footnote{This is only an approximation but it is useful for the present purpose, the two lines are:   $\Delta\fbol=1-[(\teff\pm50\mathrm{\ K})/\teff]^{2.7}$.} Note how this explains almost completely the differences found with B98.

The comparison of AAM95 fluxes with B98 may be an extreme case (see RM04b) so even if the adopted bolometric fluxes are inaccurate, a conservative estimate is that the IRFM temperatures are reliable within 50 K of systematic error. In RM04b we showed that the measurements of bolometric fluxes are within $\pm1$\% of the fluxes from AAM95 calibration, so the typical errors due to the fluxes are probably around 20 K.

\begin{figure}
\epsscale{1.1}
\plotone{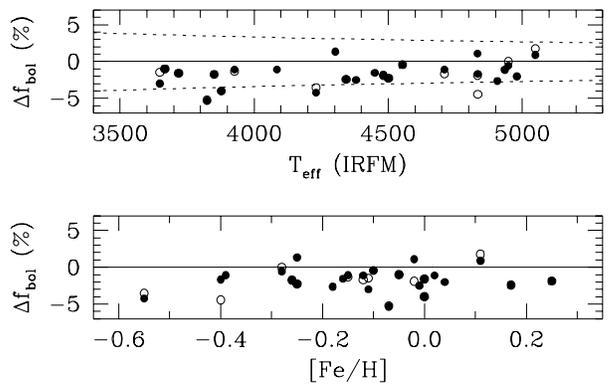} \caption{Difference between the fluxes from AAM99a calibration and the mean of the measurements found in the literature (filled circles) and M03 (open circles) as a function of $\teff$ and $\feh$. The dotted lines show how the fluxes would need to be off as to produce an IRFM $\teff$ scale shifted by $\pm$50 K.} \label{fig:fluxG}
\end{figure}

The situation with the calibration for giants (AAM99a) is similar, as it is shown in Fig.~\ref{fig:fluxG}, where the fluxes from the calibration for about 30 stars\footnote{These stars are used in \S \ref{sect:IRFMvsDIAM} to compare measured diameters and direct temperatures with the IRFM (see Table~\ref{t:theTeffG}).} are compared to the mean of the measurements we found in the literature (see \S\ref{sect:IRFMvsDIAM}) and the fluxes by Mozurkewich et~al. (2003, hereafter M03). As before, we conclude that the effect of the flux calibration on the $\teff$ scale may not be larger than 50~K of systematic error, and that typically it should be around 25~K.

\section{IRFM temperatures and diameters} \label{sect:IRFMtemps}

\subsection{Determination of $\teff$ and $\theta$}

From the definitions of the $q_\lambda$ and $R_\mathrm{theo}$ factors it is shown that
\begin{equation}
q_\lambda
R_\mathrm{theo}=\frac{\fbol}{f_\lambda^\mathrm{cal}10^{-0.4(m-m_\mathrm{cal})}}\ ,\label{eq:qR}
\end{equation}
where the left hand side is calculated from the models and filter transmission functions while the right hand side is obtained entirely from observations (adopting a calibration for the bolometric fluxes). The $qR$ factors are thus considered as the temperature indicators of the IRFM.

\begin{figure}
\epsscale{1.0}
\plotone{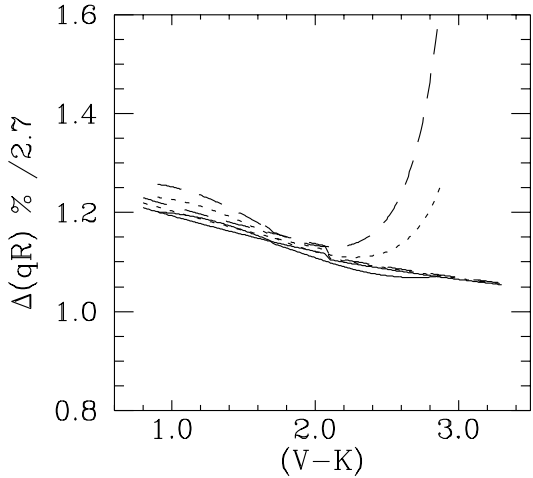} \caption{$\Delta(qR)$\% is the error in the $qR$ factors due to mean photometric uncertainties of 0.02 mag in $(V-K)$ and 0.03 mag in $(m-K)$, where $m=J,H,K,L$ (see Eq.~\ref{eq:errph}). $\Delta(qR)$\%$/2.7$ is an estimate of the corresponding error in $\teff$. Solid line: $\feh=0.0$, dotted line: $\feh=-1.0$, dashed line: $\feh=-2.0$. The set of lines in the bottom are for giant stars. The lines at top and those that diverge are for dwarfs.} \label{fig:errph}
\end{figure}

In this work, the bolometric flux calibration adopted has the functional form $\fbol=10^{-0.4K}\Phi(V-K)$. Then, from Eq. (\ref{eq:qR}) we deduce the relative error in the observed $qR$ factor:
\begin{equation}
\frac{\Delta(qR)}{(qR)}=\left\{\left[0.921\Delta(m-K)\right]^2+\left(\frac{\Delta\Phi}{\Phi}\right)^2\right\}^{1/2}\ .
\label{eq:errph}
\end{equation}
Note that $\Delta\Phi$ is a function of $(V-K)$ and its error $\Delta(V-K)$. Then, adopting mean errors of $\Delta (V-K)=0.02$ and $\Delta (m-K)=0.03$ we obtained the approximate uncertainties $\Delta(qR)$\% plotted in Fig. \ref{fig:errph}.

Since, roughly, $qR\propto T^{2.7}$ independently of the band (Fig. \ref{fig:qr}), the $\Delta(qR)\%/2.7$ given in Fig. \ref{fig:errph} is a reasonable estimate of the uncertainty in $\teff$ introduced by photometric errors. Thus, in our implementation of the IRFM we have adopted a mean error of 3\% in $qR$ as the error due to the photometry.

The errors introduced by the atmospheric parameters $\feh$ and $\logg$ were explicitly considered since their updated values are a major improvement to AAM work.

In the best cases four temperatures were derived, one for each of the bands: $J$, $H$, $K$, and $L$. Only the first three of them were available in the dwarf sample. The sensitivity of $(qR)_J$ (the $qR$ factors in the $J$ band) to $\teff$ decreases for $T<5000$ K for both dwarf and giants (Fig. \ref{fig:qr}) and thus no $T_J$ (temperature derived from the $J$ band) below 5000 K was used to get the final $\teff$. The same reason restricted the use of $T_H$ to $T_H>4200$ K for the dwarf sample. Also, not all the giant stars have $L$ magnitudes measured and so not all of them have a $T_L$.

\begin{figure*}
\epsscale{0.9}
\plotone{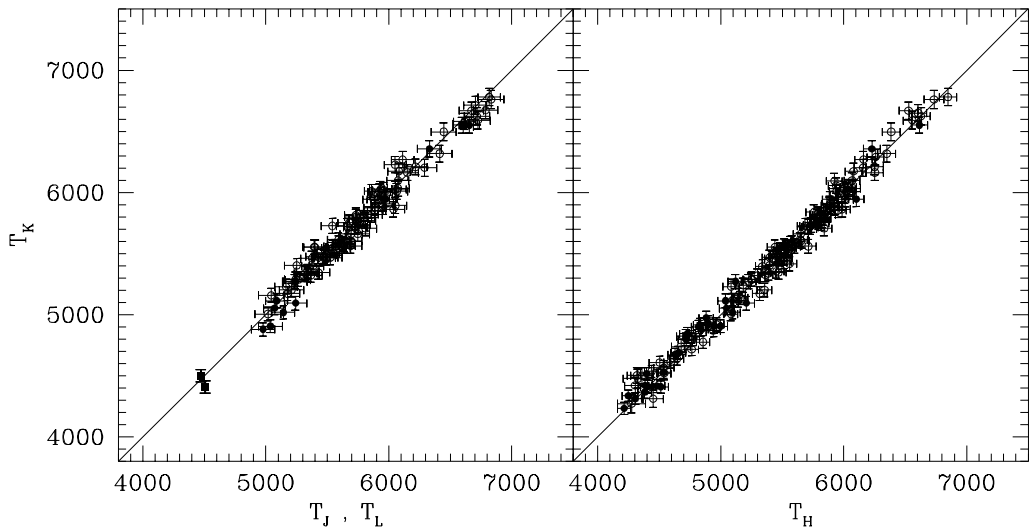} \caption{Left panel: temperatures obtained from the $J$ and $L$ bands ($T_J$, $T_L$) compared to those obtained from the $K$ band ($T_K$); open symbols: dwarfs, filled symbols: giants. Right panel: as in the left panel for $T_H$ vs. $T_K$.}
\label{fig:tjhkl}
\end{figure*}

Figure \ref{fig:tjhkl} shows that, for our sample stars, the temperatures derived in the different bands are consistent. The dispersion is larger for the dwarf sample but no trends are observed. The same result is obtained when the updated temperatures for the AAM stars are included.

The adopted IRFM effective temperature and its error was obtained from:
\begin{equation}
\teff=\frac{\sum_i[T_i(\Delta T_i)^{-1}]}{\sum_i(\Delta
T_i)^{-1}}\ \ \ \ \ , \ \ \ \ \Delta\teff=\frac{N}{\sum_i(\Delta
T_i)^{-1}}\ \ , \label{eq:teff}
\end{equation}
where $i=J,H,K,L$; and $N$ is the number of temperatures available.

The angular diameters were calculated from Eq. (\ref{eq:def1}) using the adopted bolometric fluxes (from AAM95 or AAM99a calibrations) and the IRFM temperatures. Their errors were estimated by propagating a mean error of 1.5 \% in the bolometric fluxes and the errors in $\teff$.

Table \ref{table:oursample} contains our sample stars, their adopted atmospheric parameters $\logg$, $\feh$, fluxes from AAM95 or AAM99a calibrations, reddening corrections, $(V-K)$ colors in the TCS system (transformed from either Johnson or 2MASS photometry), IRFM temperatures and their errors, and angular diameters and their errors. The table has been ordered alpha-numerically by the star standard names: HD numbers have been preferred, then BD and then G numbers. Stars with other nomenclatures have also been ordered alpha-numerically. A more detailed version of this table as well as tables for the combined sample, including the updated $\teff$ and atmospheric parameters of the AAM stars, are available upon request.

\begin{deluxetable*}{lrrcccccccccccc}
\tabletypesize{\tiny}
\tablecaption{Input Data Adopted, Effective Temperatures and Angular Diameters Derived in This Work.}
\tablehead{ID & $\logg$ & $\feh$ & $\fbol$ & $E(B-V)$ & $(V-K)$ & $\teff$ (K) & $\sigma(\teff)$ & $\theta$ (mas) & $\sigma(\theta)$}
\startdata

BD -09 1261	&	4.00	&	-0.93	&	2.034E-09	&	0.000	&	2.923	&	3967	&	148	&	0.157	&	0.012	\\
BD -10 3166	&	4.41	&	0.45	&	2.764E-09	&	0.010	&	1.953	&	5228	&	67	&	0.105	&	0.003	\\
BD -13 3442	&	3.92	&	-2.95	&	2.381E-09	&	0.000	&	1.226	&	6442	&	80	&	0.064	&	0.002	\\
BPS CS 16085-0050	&	1.00	&	-3.10	&	4.759E-10	&	0.010	&	2.225	&	4876	&	55	&	0.050	&	0.001	\\
BPS CS 22166-0030	&	4.00	&	-3.36	&	1.239E-10	&	0.033	&	1.461	&	6209	&	78	&	0.016	&	0.000	\\
\nodata	&	\nodata	&	\nodata	&	\nodata	&	\nodata	&	\nodata	&	\nodata	&	\nodata	&	\nodata	&	\nodata	\\
\enddata
\label{table:oursample}
\end{deluxetable*}

\begin{figure*}
\epsscale{0.9}
\plotone{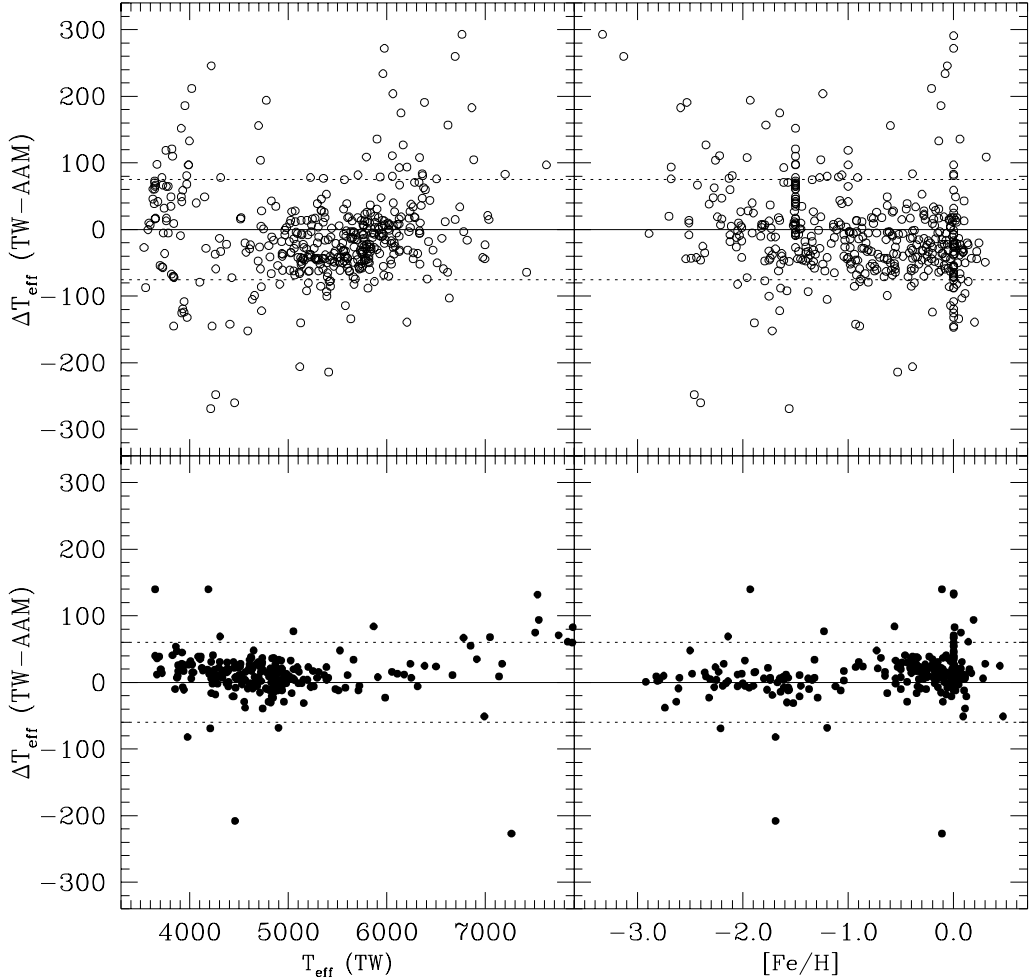} \caption{Difference between the temperatures derived in this work (TW) and those given by AAM96a and AAM99a as a function of our $\teff$ values and adopted $\feh$ for both dwarf (open circles) and giant (filled circles) stars. The dotted lines are the mean errors in our $\teff$ values (75~K for dwarf and 60~K for giant stars.).}
\label{fig:control}
\end{figure*}

\begin{figure}
\epsscale{1.0}
\plotone{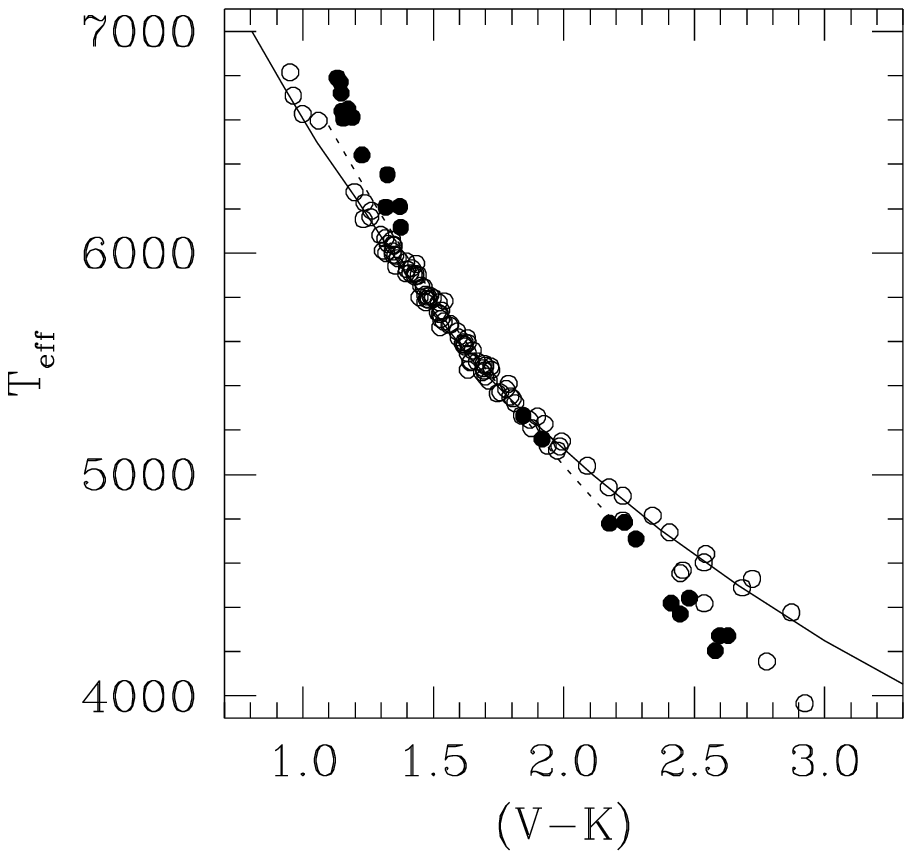} \caption{$\teff$ vs. $(V-K)$ relation for dwarfs as given by our sample stars (filled circles: $\feh<-1.5$; open circles: $\feh>-1.5$) and according to AAM96b temperature scale for $\feh=0$ (solid line) and $\feh=-3$ (dotted line).} \label{fig:vk}
\end{figure}

\subsection{The Alonso et~al. stars}

By comparing the effective temperatures we derived for the AAM sample with the AAM96a and AAM99a temperatures (Fig.~\ref{fig:control}) we find good agreement (within error bars) and thus show that the present $\teff$ scale is consistent with that by AAM within the ranges in common. A closer inspection of Fig.~\ref{fig:control} reveals very small systematic differences of about 40~K (0.7\% at solar temperature) for dwarf stars and 25~K (0.6\% at 4500~K) for giants, which are, however, well within the uncertainties of our results (1.3\%). The scatter is due to the random nature of the differences between our input data and that adopted by AAM, and the extreme cases can be easily identified by comparing the two sets of input data. Less than 9~\% (20~\%) of the giant (dwarf) stars in Fig.~\ref{fig:control} exhibit differences larger than our mean uncertainty of 60~K (75~K). If we take into account the uncertainties in AAM temperatures these numbers are even smaller.

\subsection{The need for improvement and extension}

One important improvement to the IRFM $\teff$ scale is the use of more accurate input parameters.
Even though individually the temperature differences between the temperatures we obtained and those by AAM are relatively small, overall they greately help to reduce the dispersion in the $\teff:\mathrm{color}:\feh$ relations (Part II), thus allowing to separate real effects from random errors. The revised IRFM temperature scale may then have a better \textit{internal} accuracy, which is a very useful improvement, particularly in the light of the need to reduce (or even better, recognize) the sources of uncertainty in chemical abundance studies. With the internal uncertainty reduced, the systematics can be better quantified.

Also important is the effort to include stars representative of regions of particular interest in the atmospheric parameters space, like the metal-rich (e.g. for planetary-host star studies) and metal-poor (e.g. for globular cluster and halo star studies) extremes. As an example of the importance of these selected regions we show in Fig. \ref{fig:vk} the $\teff$ vs. $(V-K)$ relation defined by our sample stars as well as AAM96b relations for $\feh=0$ and $\feh=-3$ (photometry sources and reddening corrections employed are discussed in detail in Part~II). The AAM96b relation for metal-poor cool dwarfs seems to be in agreement with ours (even if the AAM96b relation is slightly extrapolated) but the corresponding AAM96b relation for hot metal-poor stars differs from ours. The `redder' points near 6500~K on the dwarf $\teff$ vs. $(V-K)$  plane are all very metal-poor stars ($\feh\sim-3$). This is the place where our results show the most significant deviations with respect to AAM temperatures. In Part~II we show that the synthetic $\teff$ scales based on both MARCS and Kurucz models for metal-poor stars are consistent with our temperatures but not with AAM96b in this region. The inclusion of these stars allows to better define the trends of the IRFM temperature scale in that region, which has proven to be of fundamental importance, for example, in the controversy of the primordial lithium abundance (Mel\'endez \& Ram\'{\i}rez 2004).

\section{The IRFM compared with other methods} \label{sect:IRFMcomparison}

Since we have shown that our stars effective temperatures are essentially consistent with the AAM $\teff$ scale, in the comparisons below, the stars with IRFM temperatures are those for the whole sample (including the AAM stars), for which we have homogeneously calculated IRFM temperatures.

\subsection{Angular diameter measurements} \label{sect:IRFMvsDIAM}

The IRFM was explicitly developed to solve for both $\teff$ and $\theta$ simultaneously from the observed and theoretical data. One of the fundamental tests of validity of the IRFM is thus the comparison with measured angular diameters.

Richichi \& Percheron (2002, hereafter CHARM) have collected most of the published stellar angular diameters to date, measured mainly with lunar occultation (LO) and long-baseline interferometry (LBI) techniques. A large number of giant stars constitute the major part of the catalog and only a small fraction of main sequence stars is included.

\begin{figure}
\epsscale{1.1} \plotone{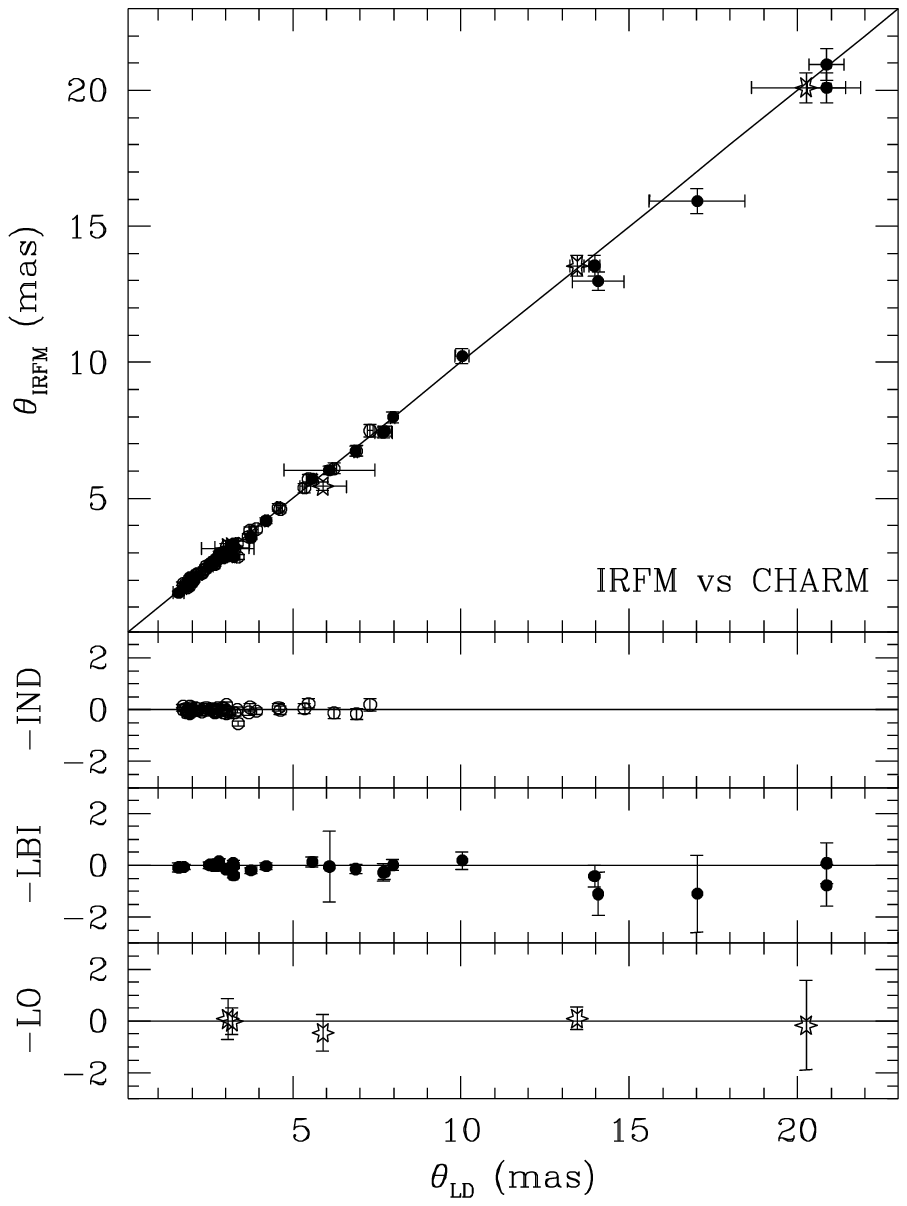}\caption{Comparison between the angular diameters from CHARM ($\theta_\mathrm{LD}$) and the diameters from the IRFM ($\theta_\mathrm{IRFM}$) for giant stars. Open circles: indirect (IND) diameters from Cohen et~al. (1999) for 55 stars; filled circles: diameters from long baseline interferometry (LBI) for 29 stars; stars: lunar occultation (LO) measurements for 5 stars. Error bars are individual uncertainties or standard deviations, the latter when more than one measurement was available. The differences between the IRFM angular diameters and the IND diameters, LBI, and LO measurements are also shown with error bars containing the uncertainties in both $\theta_\mathrm{LD}$ and $\theta_\mathrm{IRFM}$.} \label{fig:theG}
\end{figure}

We have compared the angular diameters that result from our IRFM implementation with those given in CHARM in Fig. \ref{fig:theG} (see also Table \ref{t:theTeffG}). The stars in CHARM have been grouped according to the technique employed to derive their angular diameters. First, a comparison is made with the indirect (IND) diameters of Cohen et~al. (1999), then with LBI, and finally with LO diameters.

\tabletypesize{\tiny}
\begin{deluxetable*}{rrrrrrccrrlr}
\tablecaption{\footnotesize{Giant Stars With Measured Angular Diameters in CHARM and M03}}
\tablehead{HD & $\feh$ & $\theta_\mathrm{LD}^\mathrm{CHARM}$ & \tablenotemark{1} & $\theta_\mathrm{LD}^\mathrm{M03}$ &$\teff^\mathrm{IRFM}$ &  $\fbol$\tablenotemark{2} & $\fbol$\tablenotemark{3} & $\fbol$\tablenotemark{4} & $\theta_\mathrm{IRFM}$ (mas) & $\teff^\mathrm{dir}$ \tablenotemark{5} & $\teff^\mathrm{dir}$ (M03)\tablenotemark{6}}
\startdata \tiny 
3546	&	-0.72	&	1.77	$\pm$	0.08	&	LBI	&		\nodata		&	4935	$\pm$	52	&	5.81E-07	&	\nodata	&	\nodata	&	1.71	$\pm$	0.05	&	4857	$\pm$	111	&		\nodata		\\
3627	&	0.17	&	4.21	$\pm$	0.05	&	LBI	&	4.14	$\pm$	0.04	&	4343	$\pm$	45	&	2.07E-06	&	2.12E-06	&	2.12E-06	&	4.18	$\pm$	0.11	&	4355	$\pm$	30	&	4392	$\pm$	27	\\
3712	&	-0.10	&	5.58	$\pm$	0.13	&	LBI	&	5.61	$\pm$	0.06	&	4553	$\pm$	48	&	4.68E-06	&	4.70E-06	&	4.70E-06	&	5.72	$\pm$	0.15	&	4614	$\pm$	55	&	4602	$\pm$	29	\\
6860	&	-0.07	&	14.08	$\pm$	0.76	&	LBI	&	13.75	$\pm$	0.14	&	3824	$\pm$	41	&	1.20E-05	&	1.26E-05	&	1.26E-05	&	12.98	$\pm$	0.34	&	3719	$\pm$	102	&	3763	$\pm$	23	\\
9927	&	-0.01	&	3.76	$\pm$	0.07	&	LBI	&		\nodata		&	4380	$\pm$	48	&	1.56E-06	&	1.60E-06	&	\nodata	&	3.57	$\pm$	0.09	&	4292	$\pm$	43	&		\nodata		\\
10380	&	-0.27	&	2.81	$\pm$	0.03	&	LBI	&		\nodata		&	4132	$\pm$	46	&	8.56E-07	&	\nodata	&	\nodata	&	2.97	$\pm$	0.08	&	4247	$\pm$	28	&		\nodata		\\
12533	&	-0.07	&	7.73	$\pm$	0.20	&	LBI	&	7.81	$\pm$	0.08	&	4259	$\pm$	45	&	6.12E-06	&	6.66E-06	&	6.66E-06	&	7.47	$\pm$	0.19	&	4277	$\pm$	57	&	4254	$\pm$	27	\\
12929	&	-0.25	&	6.87	$\pm$	0.04	&	LBI	&	6.83	$\pm$	0.07	&	4501	$\pm$	50	&	6.19E-06	&	6.33E-06	&	6.33E-06	&	6.73	$\pm$	0.18	&	4478	$\pm$	21	&	4493	$\pm$	28	\\
18884	&	0.00	&		\nodata		&	\nodata	&	13.24	$\pm$	0.26	&	3718	$\pm$	46	&	9.42E-06	&	\nodata	&	9.57E-06	&	12.17	$\pm$	0.35	&		\nodata		&	3578	$\pm$	37	\\
29139	&	-0.18	&	20.87	$\pm$	0.57	&	LBI	&	21.10	$\pm$	0.21	&	3883	$\pm$	44	&	3.06E-05	&	3.34E-05	&	3.33E-05	&	20.10	$\pm$	0.55	&	3895	$\pm$	56	&	3871	$\pm$	24	\\
29139	&	-0.18	&	20.26	$\pm$	1.63	&	LO	&		\nodata		&	3883	$\pm$	44	&	3.06E-05	&	3.35E-05	&	\nodata	&	20.10	$\pm$	0.55	&	3956	$\pm$	160	&		\nodata		\\
44478	&	-0.05	&	13.97	$\pm$	0.16	&	LBI	&	15.12	$\pm$	0.15	&	3667	$\pm$	44	&	1.11E-05	&	1.12E-05	&	1.12E-05	&	13.55	$\pm$	0.38	&	3623	$\pm$	25	&	3483	$\pm$	22	\\
44478	&	-0.05	&	13.45	$\pm$	0.21	&	LO	&		\nodata		&	3667	$\pm$	44	&	1.11E-05	&	\nodata	&	\nodata	&	13.55	$\pm$	0.38	&	3684	$\pm$	32	&		\nodata		\\
61338	&	0.00	&	3.07	$\pm$	0.78	&	LO	&		\nodata		&	3869	$\pm$	48	&	7.44E-07	&	\nodata	&	\nodata	&	3.16	$\pm$	0.09	&	3925	$\pm$	498	&		\nodata		\\
62509	&	-0.02	&	7.98	$\pm$	0.06	&	LBI	&	7.98	$\pm$	0.08	&	4833	$\pm$	50	&	1.16E-05	&	1.15E-05	&	1.18E-05	&	7.99	$\pm$	0.20	&	4823	$\pm$	25	&	4858	$\pm$	30	\\
62721	&	-0.27	&	3.01	$\pm$	0.00	&	LBI	&		\nodata		&	3988	$\pm$	48	&	6.94E-07	&	\nodata	&	\nodata	&	2.87	$\pm$	0.08	&	3894	$\pm$	15	&		\nodata		\\
76294	&	-0.01	&	3.24	$\pm$	0.08	&	LBI	&		\nodata		&	4817	$\pm$	50	&	1.86E-06	&	\nodata	&	\nodata	&	3.22	$\pm$	0.08	&	4806	$\pm$	61	&		\nodata		\\
80493	&	-0.26	&	7.69	$\pm$	0.26	&	LBI	&	7.54	$\pm$	0.08	&	3851	$\pm$	42	&	4.03E-06	&	4.10E-06	&	4.10E-06	&	7.42	$\pm$	0.20	&	3798	$\pm$	65	&	3836	$\pm$	24	\\
94264	&	-0.20	&	2.54	$\pm$	0.03	&	LBI	&		\nodata		&	4670	$\pm$	51	&	1.05E-06	&	\nodata	&	\nodata	&	2.58	$\pm$	0.07	&	4701	$\pm$	33	&		\nodata		\\
99998	&	-0.39	&	3.21	$\pm$	0.03	&	LBI	&		\nodata		&	3919	$\pm$	45	&	7.98E-07	&	\nodata	&	\nodata	&	3.19	$\pm$	0.09	&	3905	$\pm$	23	&		\nodata		\\
99998	&	-0.39	&	3.20	$\pm$	0.51	&	LO	&		\nodata		&	3919	$\pm$	45	&	7.98E-07	&	\nodata	&	\nodata	&	3.19	$\pm$	0.09	&	3911	$\pm$	312	&		\nodata		\\
102224	&	-0.44	&	3.23	$\pm$	0.02	&	LBI	&		\nodata		&	4378	$\pm$	46	&	1.36E-06	&	\nodata	&	\nodata	&	3.33	$\pm$	0.09	&	4447	$\pm$	22	&		\nodata		\\
112142	&	0.00	&	5.90	$\pm$	0.70	&	LO	&		\nodata		&	3647	$\pm$	47	&	1.75E-06	&	\nodata	&	\nodata	&	5.45	$\pm$	0.16	&	3504	$\pm$	208	&		\nodata		\\
113226	&	0.11	&	3.23	$\pm$	0.06	&	LBI	&	3.28	$\pm$	0.03	&	5049	$\pm$	59	&	2.25E-06	&	2.23E-06	&	2.21E-06	&	3.23	$\pm$	0.09	&	5035	$\pm$	47	&	4981	$\pm$	31	\\
124897	&	-0.55	&	20.87	$\pm$	0.52	&	LBI	&	21.37	$\pm$	0.25	&	4231	$\pm$	49	&	4.69E-05	&	4.89E-05	&	4.86E-05	&	20.95	$\pm$	0.58	&	4284	$\pm$	56	&	4226	$\pm$	29	\\
135722	&	-0.40	&	2.75	$\pm$	0.01	&	LBI	&	2.76	$\pm$	0.03	&	4834	$\pm$	50	&	1.35E-06	&	1.37E-06	&	1.41E-06	&	2.72	$\pm$	0.07	&	4831	$\pm$	20	&	4851	$\pm$	32	\\
150997	&	-0.28	&	2.52	$\pm$	0.11	&	LBI	&	2.62	$\pm$	0.03	&	4948	$\pm$	54	&	1.26E-06	&	1.27E-06	&	1.26E-06	&	2.52	$\pm$	0.07	&	4947	$\pm$	111	&	4841	$\pm$	36	\\
164058	&	-0.15	&	10.04	$\pm$	0.21	&	LBI	&	9.86	$\pm$	0.13	&	3927	$\pm$	42	&	8.29E-06	&	8.38E-06	&	8.40E-06	&	10.23	$\pm$	0.27	&	3974	$\pm$	45	&	4013	$\pm$	30	\\
189319	&	0.00	&	6.09	$\pm$	1.35	&	LBI	&	6.23	$\pm$	0.06	&	3877	$\pm$	41	&	2.75E-06	&	2.86E-06	&	2.86E-06	&	6.04	$\pm$	0.16	&	3899	$\pm$	433	&	3858	$\pm$	24	\\
197989	&	-0.12	&	5.23	$\pm$	0.60	&	LBI	&	4.61	$\pm$	0.05	&	4710	$\pm$	52	&	3.57E-06	&	3.61E-06	&	3.63E-06	&	4.67	$\pm$	0.12	&		\nodata		&	4757	$\pm$	30	\\
198149	&	-0.18	&	2.65	$\pm$	0.04	&	LBI	&		\nodata		&	4907	$\pm$	54	&	1.38E-06	&	1.42E-06	&	\nodata	&	2.67	$\pm$	0.07	&	4960	$\pm$	42	&		\nodata		\\
210745	&	0.25	&	5.31	$\pm$	0.11	&	LBI	&	5.23	$\pm$	0.05	&	4482	$\pm$	51	&	3.21E-06	&	3.27E-06	&	3.27E-06	&	4.89	$\pm$	0.13	&		\nodata		&	4351	$\pm$	27	\\
214868	&	-0.25	&	2.63	$\pm$	0.05	&	LBI	&		\nodata		&	4303	$\pm$	47	&	7.69E-07	&	7.59E-07	&	\nodata	&	2.60	$\pm$	0.07	&	4260	$\pm$	44	&		\nodata		\\
217906	&	-0.11	&	17.02	$\pm$	1.42	&	LBI	&	17.98	$\pm$	0.02	&	3648	$\pm$	43	&	1.50E-05	&	1.55E-05	&	1.52E-05	&	15.93	$\pm$	0.45	&	3557	$\pm$	149	&	3448	$\pm$	13	\\
221115	&	0.04	&	1.61	$\pm$	0.17	&	LBI	&		\nodata		&	4980	$\pm$	63	&	4.79E-07	&	4.89E-07	&	\nodata	&	1.53	$\pm$	0.05	&	4877	$\pm$	258	&		\nodata		\\
222107	&	-0.50	&	2.66	$\pm$	0.08	&	LBI	&		\nodata		&	4605	$\pm$	49	&	1.09E-06	&	\nodata	&	\nodata	&	2.70	$\pm$	0.07	&	4637	$\pm$	72	&		\nodata	
\enddata
\tablenotetext{1}{ Technique employed to derive $\theta_\mathrm{LD}$: LBI. long-baseline interferometry, LO. lunar occultation; as given in CHARM.}
\tablenotetext{2}{ Bolometric fluxes from the AAM99a calibration. Units are erg cm$^{-2}$ s$^{-1}$.}
\tablenotetext{3}{ Mean of bolometric flux measurements in erg cm$^{-2}$ s$^{-1}$. See the text for references.}
\tablenotetext{4}{ Bolometric fluxes measured by M03 in erg cm$^{-2}$ s$^{-1}$.}
\tablenotetext{5}{ Direct temperatures from the CHARM angular diameters and the mean of the bolometric flux measurements (when available) or the flux from AAM95 calibration. A mean error of 1.5\% was adopted for the fluxes.}
\tablenotetext{6}{ Direct temperatures from the M03 angular diameters and bolometric flux measurements. A mean error of 1.5\% was adopted for the fluxes.}
\label{t:theTeffG}
\end{deluxetable*}
\tabletypesize{\footnotesize}

The IND diameters are in excellent agreement with the IRFM for $\theta<8$ mas, where stars in common are found. These are diameters derived from absolutely calibrated stellar spectra in the infrared, which is the main reason for the good agreement. These are not direct measurements and will not be used to derive direct temperatures in the comparisons below.

The LBI technique uses the interference of the wavefronts from the starlight entering the telescope to resolve smaller angular sizes. For the stars with large diameters, more than one LBI entry is often found in CHARM. The mean value of those entries and its standard deviation for each star were adopted for the comparison (hence the large error bars of some points in Fig. \ref{fig:theG}), otherwise the single value and its internal error were used. Comparison with the IRFM for 29 stars shows agreement within 1.5 mas for $\theta>10$ mas and better than 0.5 mas below that value.

Less measurements with LO are available due to the limited sky coverage of the Moon ($\sim10\%$) and other observational difficulties. Nevertheless, the LO diameters are reliable and have been shown to be well scaled to the LBI ones. When compared to the IRFM, a very good agreement is also found. The differences with the five LO measurements (average in some cases) are less than 0.5~mas in the range $3<\theta<20$ mas.

\begin{figure}
\epsscale{1.1} \plotone{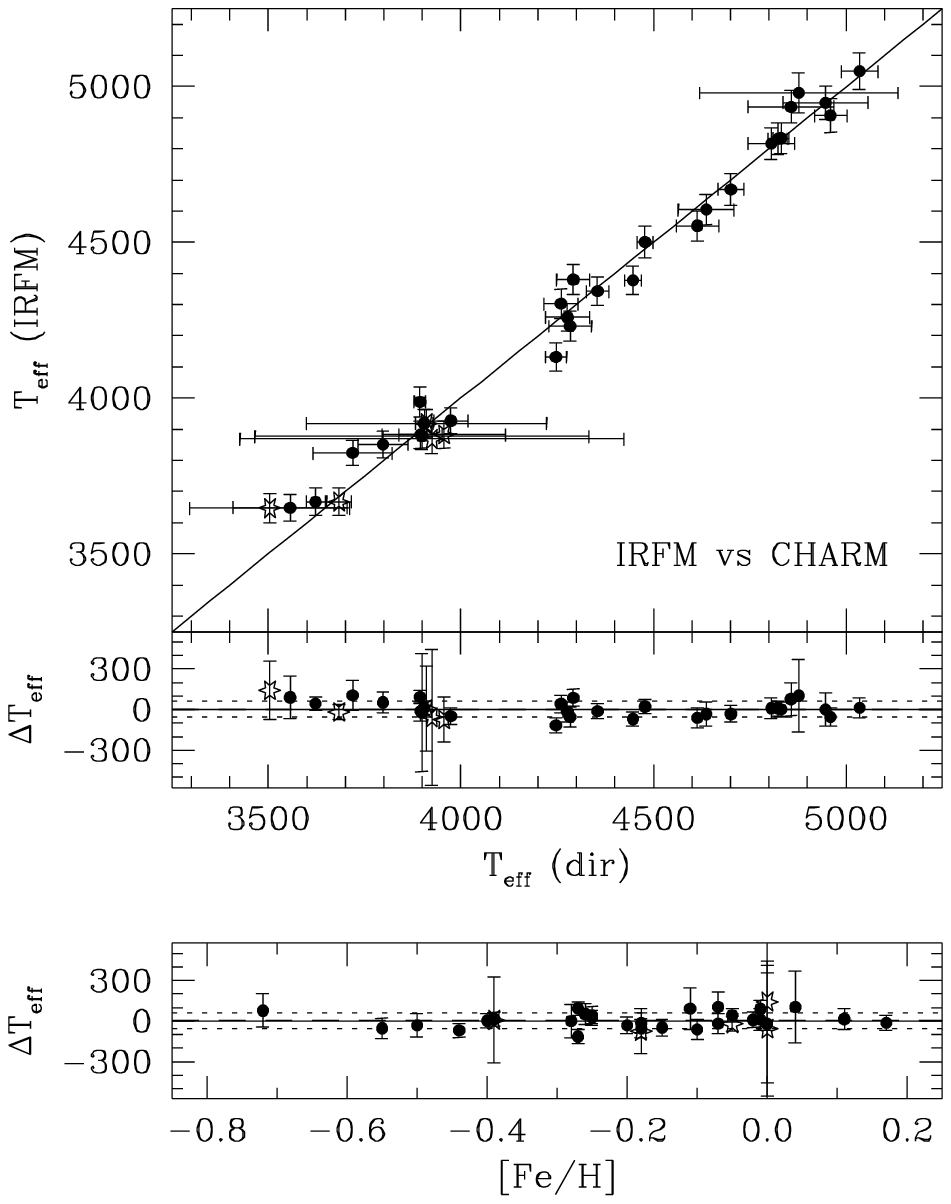} \caption{Comparison between the IRFM temperatures and the direct ones (in K) for the giants in common with CHARM. Filled circles: LBI measurements, stars: LO measurements. The differences between the IRFM temperatures and the direct ones from CHARM are also plotted as a function of $\teff$ and $\feh$ with error bars containing the uncertainties in both $\teff$ (IRFM) and $\teff$ (dir). The dashed line is at $\Delta\teff=4$ K and the dotted lines at $\Delta\teff=4\pm57$~K.} \label{fig:TeffG}
\end{figure}

From the CHARM angular diameters and the bolometric flux measurements of Blackwell et~al. (1979, 1990), Blackwell \& Lynas-Gray (1994, 1998), Leggett et~al. (1986), and M03; we obtained the direct temperatures of the stars in common with our field giant sample and compared them with those from the present IRFM implementation in Fig. \ref{fig:TeffG} (see also Table \ref{t:theTeffG}). When no flux measurements were available, the fluxes from AAM99a calibration were adopted. Although the error bars are larger than 200 K for some stars, we may safely state that a reasonable agreement is found in the ranges covered, i.e. 3500 K$<\teff<5000$ K and $-0.7<\feh<0.2$. The mean difference $(\teff^\mathrm{IRFM}-\teff^\mathrm{dir})$, with the direct temperatures from LBI measurements only, is $9\pm58$~K. The IRFM temperatures start to become larger than the direct ones at $\teff<3800$ K. Discarding the points below that limit, the difference reduces to only $4\pm57$~K. The zero point of the IRFM $\teff$ scale for giants may thus be the absolute one, but the reader should be cautioned that this result is currently only proved for stars in the ranges: 3800~K$<\teff<5000$~K and $-0.7<\feh<0.2$, where the comparison is made and no trends are found.

\begin{figure}
\epsscale{1.1} \plotone{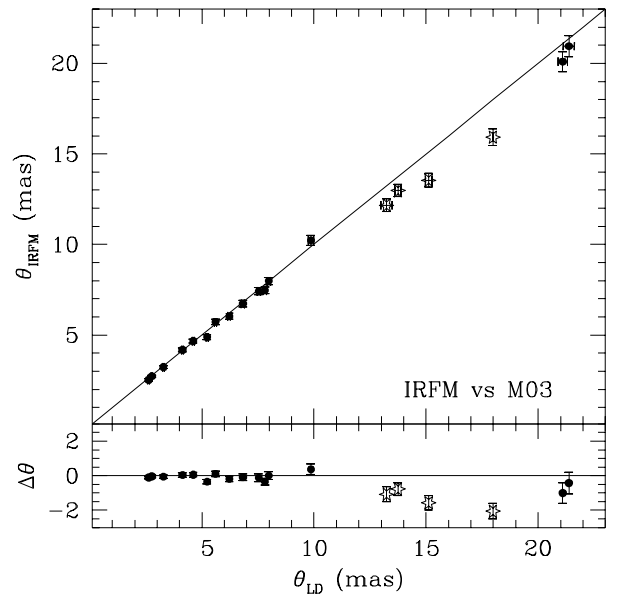}\caption{Comparison between the angular diameters from M03 ($\theta_\mathrm{LD}$) and the diameters from the IRFM ($\theta_\mathrm{IRFM}$) for 19 giant stars. Filled circles: $\teff>3800$~K, stars: $\teff<3800$~K. The differences between the IRFM angular diameters and those by M03 are also shown with error bars containing the uncertainties in both $\theta_\mathrm{LD}$ and $\theta_\mathrm{IRFM}$.} \label{fig:theGM}
\end{figure}

A more recent interferometric work is that of M03, who determined the angular diameters of 85 stars, 19 of which are included in our field giant sample. These diameters allow for a better comparison than the previous one with CHARM given the homogeneity of M03 results. The comparison between our IRFM diameters and the measurements is shown in Fig.~\ref{fig:theGM} (see also Table~\ref{t:theTeffG}). Below $\theta\sim10$ mas the agreement is excellent but above that limit the IRFM diameters appear to be lower than those by M03. The reason for this is the very low temperatures of the latter group of stars. As we stated above, the IRFM produces high temperatures compared to the direct ones below 3800 K. Thus, we have plotted the stars with $\teff<3800$ K with open symbols in Fig. \ref{fig:theGM} to show that they are the ones that depart most from the 1:1 line. Note also that M03 work uses optical interferometry and so their corrections from uniform-disk to limb-darkened diameters may be inaccurate due to uncertainties in the modeling of the TiO lines (Dyck \& Nordgren 2002).

\begin{figure}
\epsscale{1.1} \plotone{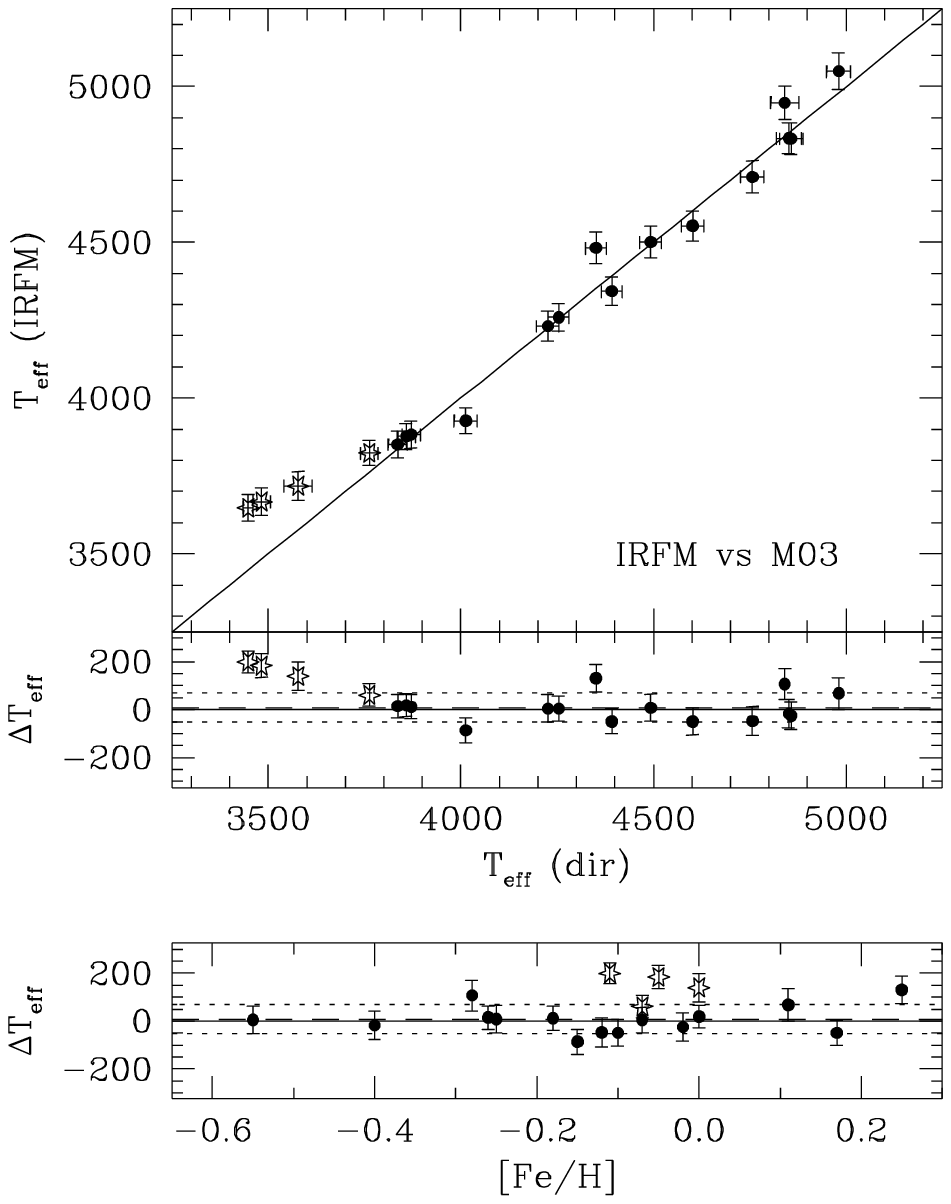} \caption{Comparison between the IRFM temperatures and the direct ones (in K) for the giants in common with M03. Filled circles: $\teff>3800$ K, stars: $\teff<3800$ K. The differences between the IRFM temperatures and the direct ones from M03 are also plotted as a function of $\teff$ and $\feh$ with error bars containing the uncertainties in both $\teff$ (IRFM) and $\teff$ (dir). The dahed line is at $\Delta\teff=8$~K and the dotted lines at $\Delta\teff=8\pm61$~K.} \label{fig:TeffGM}
\end{figure}

M03 also measured bolometric fluxes and provide direct temperatures. They are compared to the IRFM temperatures of the 19 stars in common in Fig. \ref{fig:TeffGM}. As we did before, the stars with $\teff<3800$ K have been plotted with open symbols. It is more clear now that the IRFM temperatures exceed the direct ones for the very late-type giants. Excluding them, however, the agreement is excellent, as the mean difference $(\teff^\mathrm{IRFM}-\teff^\mathrm{dir})$ is only $8\pm61$~K.

The excellent agreement found with the measured angular diameters and direct temperatures of the giant stars is not completely surprising, given that the monochromatic flux calibration in the infrared was derived in such a way that the direct temperatures of a given number of giant stars were well reproduced by the IRFM (AAM94a).

The situation for the main sequence stars is more complicated since the angular diameters of dwarf stars are very difficult to measure. Kervella et~al. (2004) have collected uniform-disk angular diameters of several dwarfs and subgiants, all measured by interferometry, and have uniformly corrected them for limb darkening.

\begin{figure}
\epsscale{1.1} \plotone{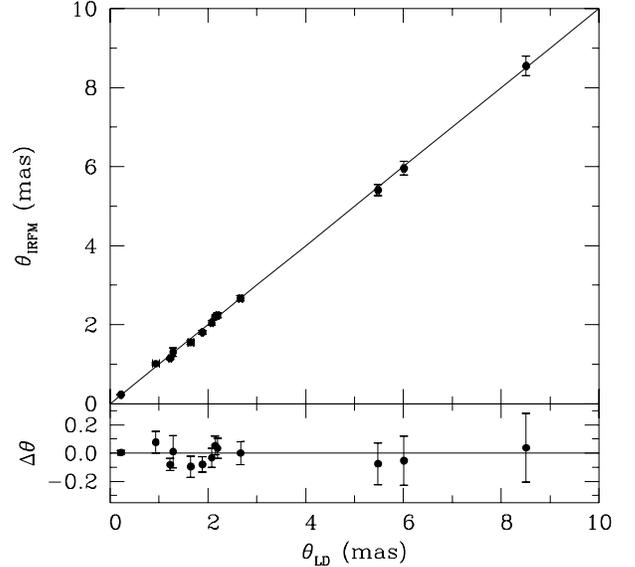} \caption{Comparison between measured angular diameters ($\theta_\mathrm{LD}$) and diameters from the IRFM ($\theta_\mathrm{IRFM}$) for the 13 dwarf (11) and subgiant (2) stars in Table \ref{t:treceD}. The differences between the IRFM angular diameters and the measurements are also shown with error bars containing the uncertainties in both $\theta_\mathrm{LD}$ and $\theta_\mathrm{IRFM}$.} \label{fig:theD1}
\end{figure}

In Fig. \ref{fig:theD1} we compare the IRFM diameters with those given by Kervella et~al. (2004) for 10 dwarf stars and 2 subgiants (see also Table \ref{t:treceD}). The two subgiants are HD 121370 and HD 198149. The original sources for the measured angular diameters include Lane et~al. (2001), Kervella et~al. (2003), Nordgren et~al. (1999, 2001), and M03; as well as some other papers in preparation (see Kervella et~al. 2004). For Procyon (HD~61421) and HD~121370 more than one value is given by Kervella et~al. (2004); only the more accurate VLTI measurement was considered. Also shown in Fig. \ref{fig:theD1} and Table \ref{t:treceD} is the comparison of the IRFM results with the transit observations of Brown et~al. (2001) for the planet-hosting star HD 209458.

\begin{deluxetable*}{rrrlccccl}
\tabletypesize{\tiny}
\tablecaption{Dwarf and Subgiant Stars With Measured Angular Diameters}
\tablehead{HD & $\feh$ & $\theta_\mathrm{LD}$ (mas) & $\teff^\mathrm{IRFM}$ (K) & $\fbol$\tablenotemark{1} & $\fbol$\tablenotemark{2} & References\tablenotemark{2} & $\theta_\mathrm{IRFM}$ (mas) & $\teff^\mathrm{dir}$ (K)}
\startdata
10700	&	-0.54	&	2.08	$\pm$	0.03	&	5372	$\pm$	65	&	1.163E-06	&	1.154E-06	&	1,2,3	&	2.047	$\pm$	0.059	&	5319	$\pm$	43	\\
16160	&	-0.03	&	0.94	$\pm$	0.07	&	4714	$\pm$	67	&	1.687E-07	&	1.725E-07	&	2,3	&	1.013	$\pm$	0.032	&	4930	$\pm$	185	\\
22049	&	-0.12	&	2.15	$\pm$	0.03	&	5015	$\pm$	56	&	1.024E-06	&	\nodata	&		&	2.204	$\pm$	0.059	&	5078	$\pm$	40	\\
26965	&	-0.28	&	1.65	$\pm$	0.06	&	5068	$\pm$	63	&	5.313E-07	&	5.313E-07	&	1,2,3	&	1.555	$\pm$	0.045	&	4920	$\pm$	91	\\
61421	&	-0.01	&	5.48	$\pm$	0.05	&	6591	$\pm$	73	&	1.837E-05	&	1.822E-05	&	3,4,5,6,7,8	&	5.405	$\pm$	0.145	&	6532	$\pm$	39	\\
88230	&	-0.12	&	1.29	$\pm$	0.04	&	3950	$\pm$	161	&	1.373E-07	&	\nodata	&		&	1.301	$\pm$	0.108	&	3967	$\pm$	63	\\
121370	&	0.25	&	2.20	$\pm$	0.03	&	6038	$\pm$	75	&	2.210E-06	&	2.206E-06	&	1,8	&	2.234	$\pm$	0.065	&	6081	$\pm$	47	\\
128620	&	0.20	&	8.51	$\pm$	0.02	&	5759	$\pm$	70	&	2.677E-05	&	\nodata	&		&	8.548	$\pm$	0.244	&	5771	$\pm$	23	\\
128621	&	0.20	&	6.01	$\pm$	0.03	&	5201	$\pm$	65	&	8.653E-06	&	\nodata	&		&	5.957	$\pm$	0.173	&	5178	$\pm$	23	\\
131977	&	0.09	&	1.23	$\pm$	0.03	&	4571	$\pm$	52	&	1.921E-07	&	2.010E-07	&	2	&	1.149	$\pm$	0.031	&	4469	$\pm$	57	\\
198149	&	-0.18	&	2.67	$\pm$	0.04	&	4907	$\pm$	54	&	1.377E-06	&	1.414E-06	&	1	&	2.670	$\pm$	0.071	&	4939	$\pm$	41	\\
209100	&	-0.02	&	1.89	$\pm$	0.02	&	4642	$\pm$	54	&	5.069E-07	&	5.000E-07	&	2	&	1.810	$\pm$	0.050	&	4527	$\pm$	29	\\
209458	&	-0.01	&	0.23	$\pm$	0.02	&	5993	$\pm$	71	&	2.279E-08	&	\nodata	&	\nodata	&	0.230	$\pm$	0.006	&	6049	$\pm$	202
\enddata
\tablenotetext{1}{ Bolometric fluxes from the AAM95 calibration. Units are erg cm$^{-2}$ s$^{-1}$.}
\tablenotetext{2}{Mean of bolometric flux measurements in erg cm$^{-2}$ s$^{-1}$. References: 1. Blackwell \& Lynas-Gray (1994), 2. Blackwell \& Lynas-Gray (1998), 3. AAM95, 4. Code et~al. (1976), 5. Beeckmans (1977), 6. Blackwell \& Shallis (1977), 7. Smalley \& Dworetsky (1995), 8. M03.}
\label{t:treceD}
\end{deluxetable*}

\begin{figure}
\epsscale{1.1} \plotone{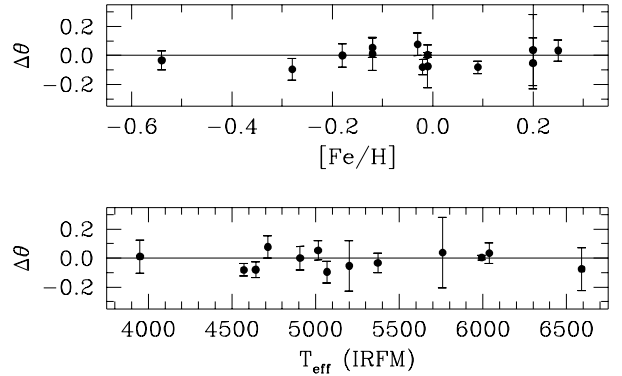} \caption{Difference between the IRFM angular diameters and the measurements as a function of the metallicity and the IRFM effective temperature for the 13 dwarf (11) and subgiant (2) stars in Table \ref{t:treceD}. Units of $\Delta\theta$ are mas. The error bars contain the uncertainties in both $\theta_\mathrm{LD}$ and $\theta_\mathrm{IRFM}$.} \label{fig:theD2}
\end{figure}

The comparison shows that the IRFM diameters are in excellent agreement with the measurements, to better than 0.2 mas in $\theta$. Also, this result seems to be valid for the wide temperature range from 4000~K to 6500~K, although it is restricted to almost solar metallicity (Fig.~\ref{fig:theD2}). The stars in Table \ref{t:treceD} span the metallicity range from $-0.54$ to 0.25 dex.

Table \ref{t:treceD} also contains the bolometric fluxes from AAM95 calibration and those measured by several authors. When a measured $\fbol$ was available, the direct temperature was derived from it and the measured angular diameter using Eq. (\ref{eq:def1}), otherwise the flux from AAM95 calibration was adopted. The mean of the measurements of $\fbol$ is given in Table \ref{t:treceD}. The direct temperatures and their errors are also given in Table \ref{t:treceD}, a mean error of 1.5\% in $\fbol$ was assumed.

\begin{figure}
\epsscale{1.1} \plotone{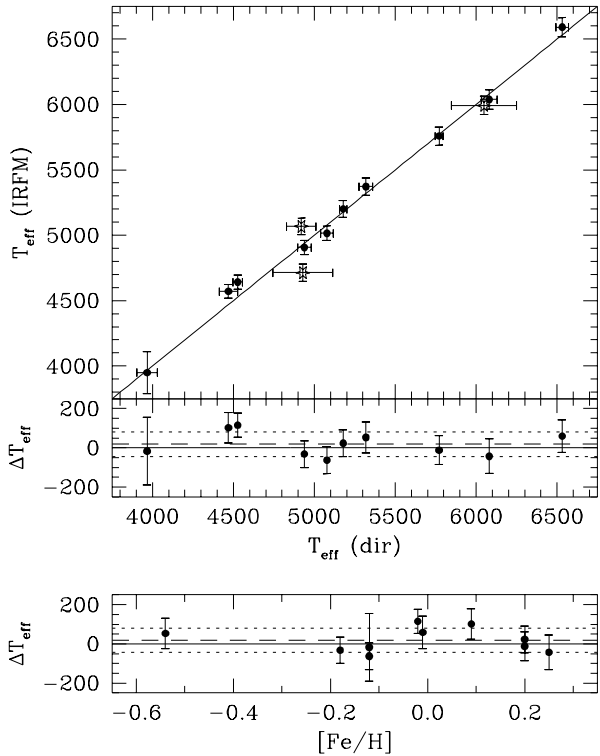} \caption{Comparison between the IRFM temperatures and the direct ones (in K) for the 13 dwarf (11) and subgiant (2) stars in Table~\ref{t:treceD}. The three open symbols are HD~16160, HD~26965, and HD~209458; they are not plotted in the residuals. The differences between the IRFM temperatures and the direct ones as a function of $\teff$ and $\feh$ are also shown with error bars containing the uncertainties in both $\teff$ (IRFM) and $\teff$ (dir). The dahed line is at $\Delta\teff=18$ K and the dotted lines at $\Delta\teff=18\pm62$~K.} \label{fig:TeffD}
\end{figure}

The comparison between direct and IRFM temperatures is shown in Fig. \ref{fig:TeffD}. The three points that depart most from the 1:1 line are HD 16160, HD 26965 and HD 209458. The uncertainty in the angular diameters measured for these stars is larger than 4\% and, thus, they are not useful to constraint any $\teff$ scale within 200 K. Excluding these three points, the mean difference $\teff^\mathrm{IRFM}-\teff^\mathrm{dir}$ is $18\pm62$~K. Consequently, the zero point of the IRFM $\teff$ scale for dwarfs is also in good agreement with the direct $\teff$ scale, given that the $18\pm62$~K difference is well within the average uncertainty of the IRFM temperatures ($\sim75$ K), at least for stars with temperatures between 4000 K and 6500 K and metallicities in the range $-0.55<\feh<0.25$.

\subsection{The Sun, Procyon, and Arcturus}

The temperatures of these three stars are very well documented, although no general agreement concerning their exact values (the Sun being the exception) has been achieved, and should be used to test any $\teff$ scale.

Using the photometry given by AAM94a, the present IRFM temperature of the Sun results in $5756\pm71$~K which is 21~K lower than the standard value. AAM96a obtained $5763\pm90$~K and thus the difference with our solar $\teff$ may only be attributed to the details of the calculations. The fact that the IRFM does not exactly reproduce the solar temperature has been suggested to imply that the zero point of the IRFM $\teff$ scale is off by some amount. This is by no means conclusive, as the photometry of the Sun is very uncertain and the errors in the observed magnitudes and colors are easily propagated to the temperature derived with this method.

Procyon (HD 61421), whose spectral type, as given in SIMBAD, is F5IV-V, has an IRFM temperature of $6591\pm73$~K. AAM96a obtained $6579\pm100$~K, the difference is mainly due to a slightly different set of $\feh$ and $\logg$ values adopted. Its direct temperature is $6532\pm39$~K, as given in Table~\ref{t:treceD}. These different results may be consistent, considering the size of the error bars. Being a bright star, the photometry has a considerable uncertainty and may be the reason for the discrepancy, if any. Using the spectroscopic method, Takeda et~al. (2002b, hereafter T02b) obtained 6576 K; Edvardsson et~al. (1993, hereafter E93), on the other hand, obtained 6704 K from a theoretical calibration of the $(b-y)$ color.  A fit to spectrophotometric data by Castelli et~al. (1997) resulted in $\teff=6650$~K. Fitting of the Balmer line-profiles in the spectrum of Procyon leads to $6667$~K, $6470$~K, and $6507$~K according to Castelli et al. (1997), F98 and AP04, respectively.

For the giant star Arcturus (HD 124897), the present IRFM implementation gives $4231\pm49$~K. That of AAM99a was $4233\pm55$~K. Its direct temperature, as given by Griffin \& Lynas-Gray (1999) is $4290\pm28$~K; the value we derived for this star in Table \ref{t:theTeffG} from several angular diameter and bolometric flux measurements is $4284\pm56$~K, in perfect agreement with Griffin \& Lynas-Gray result. M03 also derived a direct temperature for Arcturus, which is slightly lower (4226~K), but in better agreement with the IRFM.

Although the IRFM temperature of Arcturus is lower by about 60 K, the temperature derived from several temperature-color calibrations based on the IRFM is closer to the direct one (4296 K, RM04a). Thus, the use of several of these calibrations allows to get rid of errors in single IRFM temperatures, which are mainly based on infrared magnitudes alone. We will come back to this point in Part II of this work.

\subsection{Other semi-direct methods}

In \S\ref{sect:intro} we discussed some of the other methods developed to derive $\teff$. As it is the case with the IRFM, they all introduce models to complement the available observations and so the differences that we will describe below are partly due to the different models adopted, although the observational input parameters may also be the reason of the discrepancies, if any.

\begin{figure}
\epsscale{1.1} \plotone{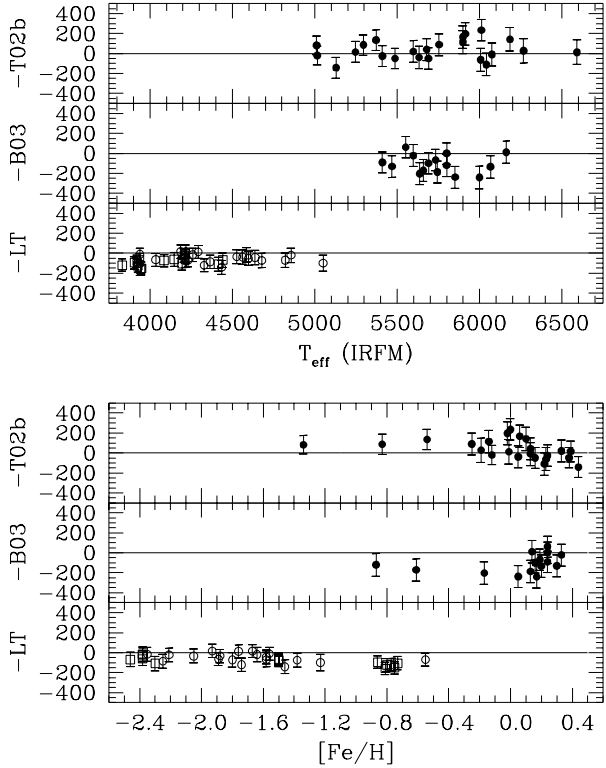} \caption{Comparison of the IRFM temperatures with those derived by other authors. The figures show the IRFM temperatures minus those given by T02b, B03, and LT as a function of the IRFM $\teff$ and our adopted values of $\feh$. Filled circles: dwarfs; open circles: field giants; open squares: giants in clusters.} \label{fig:compSP}
\end{figure}

The comparison with the spectroscopic temperatures derived by three different groups is given in Fig.~\ref{fig:compSP}. In this and the next two figures the filled circles are dwarf stars, the open circles are field giants and the open squares are giants in clusters.

T02b applied spectroscopic equilibrium to obtain the temperatures of 32 F-K dwarfs and subgiants, 23 of which are included in our dwarf sample. The comparison shows that the IRFM temperatures are higher by about 35 K but the dispersion is large ($\sim 100$ K), mainly due to a metallicity effect. Below solar metallicity, T02b temperatures are systematically lower; excluding the stars with $\feh<0$, the mean difference reduces to 10~K, but then a clear trend with temperature is observed.

Comparison with B03 spectroscopic temperatures of a sample of thin and thick disk stars suggests a mean difference of $-110\pm90$~K, in which the dispersion is, again, mainly due to a trend of the differences with metallicity. It seems that the differences reduce to zero at $\feh\sim0.25$ and the B03 temperatures become systematically higher by 110 K as lower metallicities are reached. A similar difference was found in RM04b between the IRFM temperatures and the spectroscopic ones of SIM04 for the planet-hosting stars but that one did not show any trend with metallicity, and only a slight trend with temperature, probably due to non-LTE effects and/or the bolometric fluxes adopted for the IRFM.

In Fig.~\ref{fig:compSP} is also shown a comparison of the IRFM temperatures of 21 metal-poor field giants ($-2.74<\feh<-0.55$) and 14 giants in the globular clusters M3 ($\feh\sim-1.5$), M71 ($\feh\sim-0.8$) and M92 ($\feh\sim-2.4$) with those derived spectroscopically by Sneden et~al. (1991), Kraft et~al. (1992), and Shetrone (1996), which we refer jointly as LT (Lick-Texas group). The differences have a mean value of $-65\pm50$ K, they show no large trends with $\teff$ nor $\feh$, and it is also independent of the population, i.e. it is basically the same for the giants in the field and in the three clusters. Thus, besides an offset of 65 K, the spectroscopic $\teff$ scale of the metal-poor $(-2.5<\feh<-0.5)$ giants is differentially consistent with the IRFM.

\begin{figure}
\epsscale{1.1} \plotone{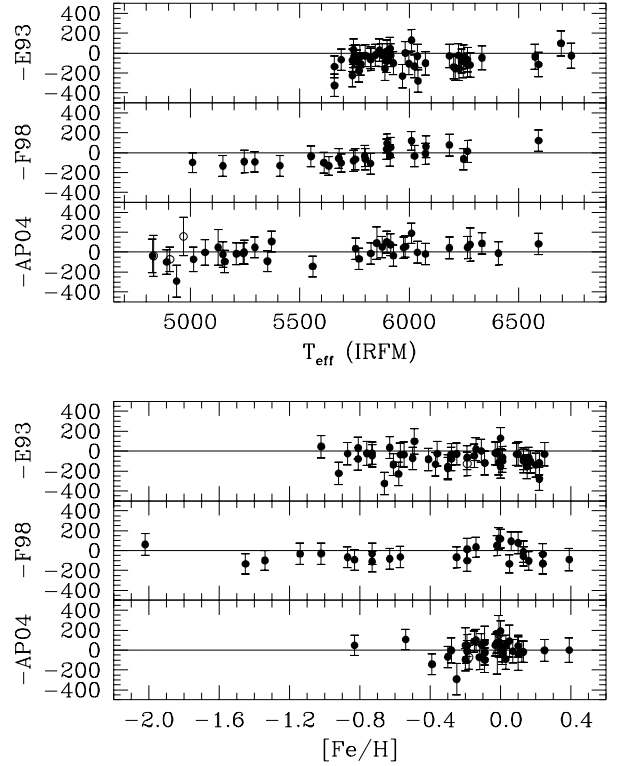} \caption{As in Fig. \ref{fig:compSP} for E93, F98, and AP04.} \label{fig:compEB}
\end{figure}

Using spectra synthesized from MARCS models, the filter transmission functions of the $uvby$-$\beta$ system, and a set of stars to fix the zero point; E93 calculated a grid of synthetic colors from which the temperatures of their sample stars were determined. We found 50 stars (1 giant) in common with our sample. The comparison with the IRFM is shown in Fig.~\ref{fig:compEB} where a mean difference of $-75\pm85$ K is observed. No trend is apparent with metallicity, and probably a slight trend with temperature is present, in the sense that the IRFM temperatures and those by E93 may be in agreement for $\teff>6500$ K but with a decreasing difference for lower temperatures.

Two sets of temperatures based on the fitting of the Balmer line-profiles are shown in Fig.~\ref{fig:compEB}. Note how even though F98 and AP04 temperatures were derived using the same principle, the theoretical prescriptions adopted have an important effect on the final results. F98 temperatures are higher by about 100 K below 5800 K but slightly lower than the IRFM ones above 6000 K. There is apparent agreement at solar metallicity but disagreement both above and below $\feh\sim0$. The solar temperature is thus well reproduced by both the IRFM and F98. On the other hand, AP04 temperatures show very good agreement with ours in the wide temperature range from 4800 K to 6600 K and it is likely that they are close to the IRFM ones even at low metallicites ($\feh\sim-0.8$).

Note that the zero point of the E93 theoretical calibration was set using a sample of stars with temperatures derived from Balmer line-profile fitting (Cayrel de Strobel 1990) and so the similarity between the differences found with E93 and F98 may be due to the same basic technique employed. We recall, however, that this technique is very model dependent, as shown by the most recent and improved results of AP04, and that it is very likely that the results from the IRFM are in good agreement with those from the modeling of hydrogen lines.

\begin{figure}
\epsscale{1.1} \plotone{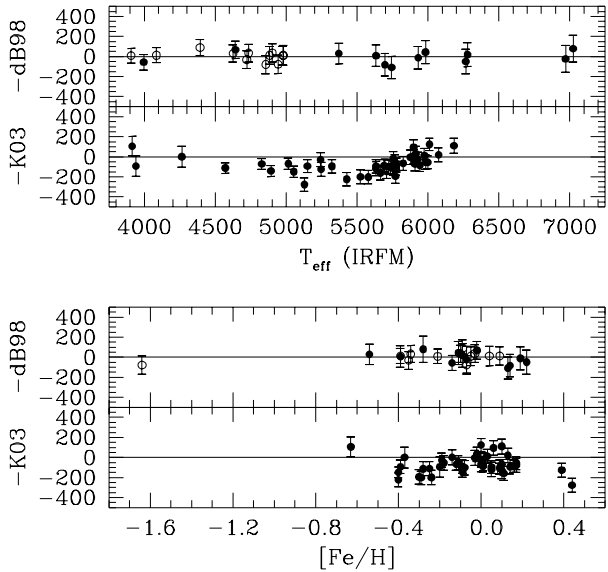} \caption{As in Fig. \ref{fig:compSP} for dB98 and K03.} \label{fig:compDiK}
\end{figure}

A method that was not discussed in \S\ref{sect:intro} is the one that uses surface brightness vs. color relations based on measurements of stellar angular diameters (di Benedetto 1998, hereafter dB98). Since the IRFM is also based on measured angular diameters, the zero points of both the IRFM and the dB98 $\teff$ scales should be the same. This is actually shown in Fig. \ref{fig:compDiK}, where the temperatures of 25 stars (12 giants) in common with our sample are compared to those given by dB98. The mean difference is $-1\pm53$ K, as expected, and the dispersion is fully consistent with the mean errors of the IRFM and dB98 temperatures. The good agreement of the zero point is also true if dwarfs and giants are considered separately. Note, however, that the comparison with dB98 results is limited to $\feh>-0.6$, although the only point below $-0.6$ dex, a giant star with $\feh=-1.64$, shows a dB98 temperature consistent with the IRFM. 

Finally, the line-depth ratio technique has been applied to 181 dwarfs by K03, 44 of them are included in our sample. It is claimed that very small errors, of the order of 10 K or less, may be achieved with this technique if several calibrated ratios are used. The comparison with our IRFM temperatures, however, shows a dispersion that is larger than the one expected by the combination of the errors in the IRFM temperatures ($\sim75$~K) with those suggested by K03, which will be essentially equal to the uncertainty from the IRFM. It is true that very small \textit{individual} uncertainties (that is, for a single star) may be achieved with the line-depth ratio technique, but some systematic effects may enlarge the dispersion in the $\teff$ scale derived with this technique, which is, however, potentially useful for studying intrinsic temperature fluctuations. K03 adjusted their $\teff$ scale to reproduce the solar temperature, hence the good agreement with the IRFM around 5800~K and solar metallicity. From 5500~K to 6200~K, however, K03 temperatures become lower than the IRFM ones. From 4500~K to 5500~K, a mean difference of about $-90$~K is observed, and it seems to go to zero again at lower temperatures.

\section{Conclusions} \label{sect:conclusions}

Effective temperatures with a mean accuracy of about 1.3\% have been determined with the infrared flux method for a selected sample of stars as well as for the Alonso et~al. (AAM) sample with the aim of calibrating the metallicity-dependent temperature vs. color relations in a companion publication. The status of this particular IRFM implementation is revised and it is shown that the IRFM temperatures are well scaled to the direct ones for main sequence stars. The case for giant stars was proved in the past and the present work confirms the good agreement. The present work improves the IRFM $\teff$ scale by the use of up-to-date spectroscopic atmospheric parameters and the inclusion of stars in the ranges where few or no data were previously studied. Our IRFM approach, however, is largely based on the AAM study, and a comparison of our updated to AAM temperatures shows agreement within error bars.

We point out that the influence of the bolometric flux calibrations adopted on our IRFM temperatures is important because they may introduce systematic trends in the resulting $\teff$ scale. However, we have also shown that the systematic trends may not be larger than 50~K, and so, if differences larger than 50~K are found with other $\teff$ scales, they should not be attributed solely to the fluxes adopted in the IRFM. This number (50~K) is a conservative estimate, and thus even smaller differences may not be totally due to errors in the bolometric fluxes.

Comparison of the stellar angular diameters from the IRFM with those measured by interferometry, lunar occultation or transit observations shows good agreement. This work thus provides accurate stellar angular diameters, which are essential to calibrate extremely high-resolution long-baseline interferometric observations (e.g. VLT and CHARA interferometers).

When combined with bolometric flux measurements, the interferometric angular diameters produce a temperature scale with a zero point that is only 10~K below that derived with the IRFM for the giant stars with $\feh>-0.7$ and temperatures above 3800~K and below 5000~K. The corresponding zero points of the dwarf $\teff$ scales differ by only about 20~K (the IRFM being larger), and it is valid for $\feh>-0.55$ and temperatures between 4000~K and 6500~K. No stars outside those ranges were available for comparison, and we cannot guarantee that the zero point of the IRFM $\teff$ scale is essentially the absolute one everywhere in the atmospheric parameters space, as it is the case in the ranges mentioned above.

The IRFM temperature of the Sun is only about 20~K below the standard value, and those of the well studied stars Procyon and Arcturus are consistent with the direct ones and those derived with other methods within the error bars. No perfect agreement is expected for these bright stars because the photometric uncertainties associated are large and easily propagated in the IRFM. Consequently, no empirical shifts to reproduce the temperature of the Sun or other well studied stars were applied.

The comparison with the temperatures obtained with other methods shows that the temperature scale of the lower main sequence is still unreliable at the 100 K level because larger discrepancies are found. Although consistent with the temperatures from Balmer line-profile fitting, the IRFM produces temperatures that are about 100 K lower that those derived from the excitation equilibrium of \ion{Fe}{1}, but exceptions are found. The temperature scale of giant stars is better defined, probably to a 50~K level or even better. The spectroscopic temperatures of the metal-poor giants, both in the field and in clusters, are about 65~K higher that those from the IRFM, but otherwise differentially consistent with them.

With the IRFM temperatures described in this paper and the atmospheric parameters adopted, we will calibrate a set of homogeneous temperature vs. color relations (Part II). The fact that the input parameters are somewhat inhomogeneous (several sources have been considered) will not have a significant effect on these relations but they may have some effect on the IRFM temperatures.

\acknowledgments{We thank A. Alonso for early discussions on this work and for providing some of his published data in electronic format, and P. S. Barklem for providing the temperatures from Balmer lines used in AP04. JM thanks partial support from NSF grant AST-0205951 to J.~G.~Cohen. This publication makes use of data products from 2MASS, which is a joint project of the University of Massachusetts and IPAC/Caltech, funded by NASA and the National Science Foundation; and the SIMBAD database, operated at CDS, Strasbourg, France.}

\end{document}